\documentclass[manuscript]{aastex}
\usepackage{natbib,graphicx,amsmath,amssymb,bm,multirow,yfonts}

\newcommand{\be}{\begin{equation}}
\newcommand{\ee}{\end{equation}}
\newcommand{\bea}{\begin{eqnarray}}
\newcommand{\eea}{\end{eqnarray}}
\newcommand{\ba}{\begin{align}}
\newcommand{\ea}{\end{align}}

\newcommand{\gcmiq}{\, \text{g} \, \text{cm}^{-3}}
\newcommand{\mev}{\, \text{MeV}}

\newcommand{\om}{\omega}
\newcommand{\qq}{{\bf q}}

\newcommand{\pp}{{\bf p}}
\newcommand{\PP}{{\bf P}}
\newcommand{\kk}{{\bf k}}

\newcommand{\rme}{{\rm e}}
\newcommand{\rmi}{{\rm i}}

\begin{document}

\title{Neutrino processes in partially degenerate neutron matter}

\author{S.\ Bacca\altaffilmark{1}}
\author{K.\ Hally\altaffilmark{1,2}}
\author{M.\ Liebend\"orfer\altaffilmark{3}}
\author{A.\ Perego\altaffilmark{3}}
\author{C.\ J.\ Pethick\altaffilmark{4,5}}
\author{A.\ Schwenk\altaffilmark{6,7}}

\altaffiltext{1}{TRIUMF, 4004 Wesbrook Mall, Vancouver, BC, V6T 2A3, Canada}
\altaffiltext{2}{Ottawa-Carleton Institute for Physics, Carleton University, 
Ottawa, K1S 5B6, Canada}
\altaffiltext{3}{Department of Physics, University of Basel, 
Klingelbergstr. 82, 4056 Basel, Switzerland}
\altaffiltext{4}{The Niels Bohr International Academy, 
The Niels Bohr Institute, Blegdamsvej 17, \mbox{DK-2100} Copenhagen \O, Denmark}
\altaffiltext{5}{NORDITA, Royal Institute of Technology and Stockholm
University, Roslagstullsbacken~23, SE-10691 Stockholm, Sweden}
\altaffiltext{6}{ExtreMe Matter Institute EMMI, GSI Helmholtzzentrum f\"ur
Schwerionenforschung GmbH, D-64291 Darmstadt, Germany}
\altaffiltext{7}{Institut f\"ur Kernphysik, Technische Universit\"at
Darmstadt, D-64289 Darmstadt, \mbox{Germany}}

\begin{abstract}
We investigate neutrino processes for conditions reached in
simulations of core-collapse supernovae. Where neutrino-matter
interactions play an important role, matter is partially degenerate,
and we extend earlier work that addressed the degenerate regime.
We derive expressions for the spin structure factor in neutron 
matter, which is a key quantity required for evaluating rates of
neutrino processes. We show that, for essentially all conditions
encountered in the post-bounce phase of core-collapse supernovae,
it is a very good approximation to calculate the spin relaxation
rates in the nondegenerate limit. We calculate spin relaxation rates
based on chiral effective field theory interactions and find that 
they are typically a factor of two smaller than those obtained 
using the standard one-pion-exchange interaction alone.
\end{abstract}


\maketitle

\section{Introduction}

Understanding the mechanisms responsible for supernovae and neutron
star formation requires a knowledge of the equation of state and
transport mechanisms in matter at densities of the order of that in
atomic nuclei and at temperatures ranging up to $10^{11}$K.  Despite
almost half a century of work on the subject, the question of how a
fraction of the large thermal energy from the collapse of the core is
transferred to the outer stellar layers, thereby causing a supernova
explosion, is one that has yet to find a convincing answer.  Because
of the high matter density in the new-born protoneutron star, most
macroscopic transport processes are ineffective, and a variety of
other mechanisms have been considered. These include energy transfer
by neutrinos that interact with matter via weak interactions
\citep{Colgate.White:1966,Bethe.Wilson:1985}, magnetic fields in
combination with rotation
\citep{Ardeljan.Bisnovatyi.ea:2004,Kotake.Sato.Takahashi:2006},
convection \citep{Herant.Benz.ea:1994}, or pressure waves and Alfv\'en
waves that may be emitted by the protoneutron star
\citep{Burrows.Livne.ea:2007,Sagert.Fischer.ea:2009}.  Some of these
mechanisms lead to very characteristic features in the gravitational
wave \citep{Ott:2009} and neutrino signatures
\citep{Dasgupta.Fischer.ea:2010}, and these will be constrained by
observations of the next galactic supernova. Other mechanisms have
been tested in simulations, which demonstrate that, with current
microphysical input, neutrino transport alone is insufficient to
generate explosions \citep{Liebendoerfer.Mezzacappa.ea:2001,%
Rampp.Janka:2002,Thompson.Burrows.Pinto:2003}. However, it is
possible to obtain explosions when neutrino heating and fluid
instabilities are combined in axisymmetric models of stellar core
collapse \citep{Buras.Rampp.ea:2006,Marek.Janka:2009,Bruenn2009,%
Suwa:2009py,Brandt:2010xa}.

In all the above mechanisms, transport by neutrinos is crucial: it is
the dominant process for energy emission, it contributes to the
transport of energy and lepton number, it influences the radial
entropy and lepton fraction gradients that determine the stellar
structure and fluid instabilities, and it gives rise to the neutrino
luminosities and spectra, which will become the main observables for
probing the properties of matter at high density in the next nearby
supernova event.  Calculations of rates of neutrino processes in dense
matter shortly after the discovery of weak neutral currents were
reviewed by \citet{Freedman}.  In the early work, the particles
participating in the reactions were taken to be free, but,
subsequently, effects of strong interactions were taken into account
\citep{sawyer, Iwamoto}. More recently, detailed calculations of rates
of neutrino processes have been performed within a mean-field approach
(the random phase approximation) by \citet{Reddy,BurrowsSawyer,Reddy2,%
BurrowsReddyThompson}. One effect not included in these calculations
is that excitations can decay due to interactions in the dense medium.
As stressed in \citet{RaffeltSeckel,RaffeltSeckelSigl,HR}, this can
lead to an energy transfer in neutrino processes considerably greater
than that predicted on the basis of excitations with infinitely long
lifetimes (i.e., from recoil effects alone). \citet{LOP} showed how to
include these damping effects in a mean-field approach, and a unified
approach to structure factors was described by \citet{LPS}.
Detailed calculations were performed based on chiral effective field
theory (EFT) interactions in \citet{Bacca} for degenerate neutrons. The
prime purpose of the present paper is to extend these studies to
partially degenerate and nondegenerate conditions.

The plan of the paper is as follows. In Section~\ref{conditions}, we
analyze the results of simulations of core-collapse supernovae and
determine the conditions for which it is important to know rates of
neutrino processes.  These results point clearly to the need for a
better understanding of neutrino properties in regions where nucleons
are partially degenerate or nondegenerate. In Section~\ref{formalism},
we develop a general formalism based on Landau's theory of normal
Fermi liquids for calculating structure factors of strongly
interacting matter. The spin relaxation rate for partially degenerate
conditions is derived in Section~\ref{time} and the nondegenerate
limit is studied in Section~\ref{nondegen}. A key
quantity is the spin relaxation rate in partially degenerate neutron
matter, and we calculate this in Section~\ref{results} based on the
one-pion-exchange approximation for nucleon-nucleon interactions,
which is the standard one used in simulations \citep{HR}, and from chiral
EFT interactions. Particular attention is paid to conditions of
importance in supernova simulations. Finally, we summarize and give
future perspectives in Section~\ref{summary}.

\section{Physical conditions in stellar collapse}
\label{conditions}

\begin{figure*}[t]
\begin{center}
\includegraphics[trim=3mm 4mm 14mm 116mm,clip,scale=0.7,clip=]{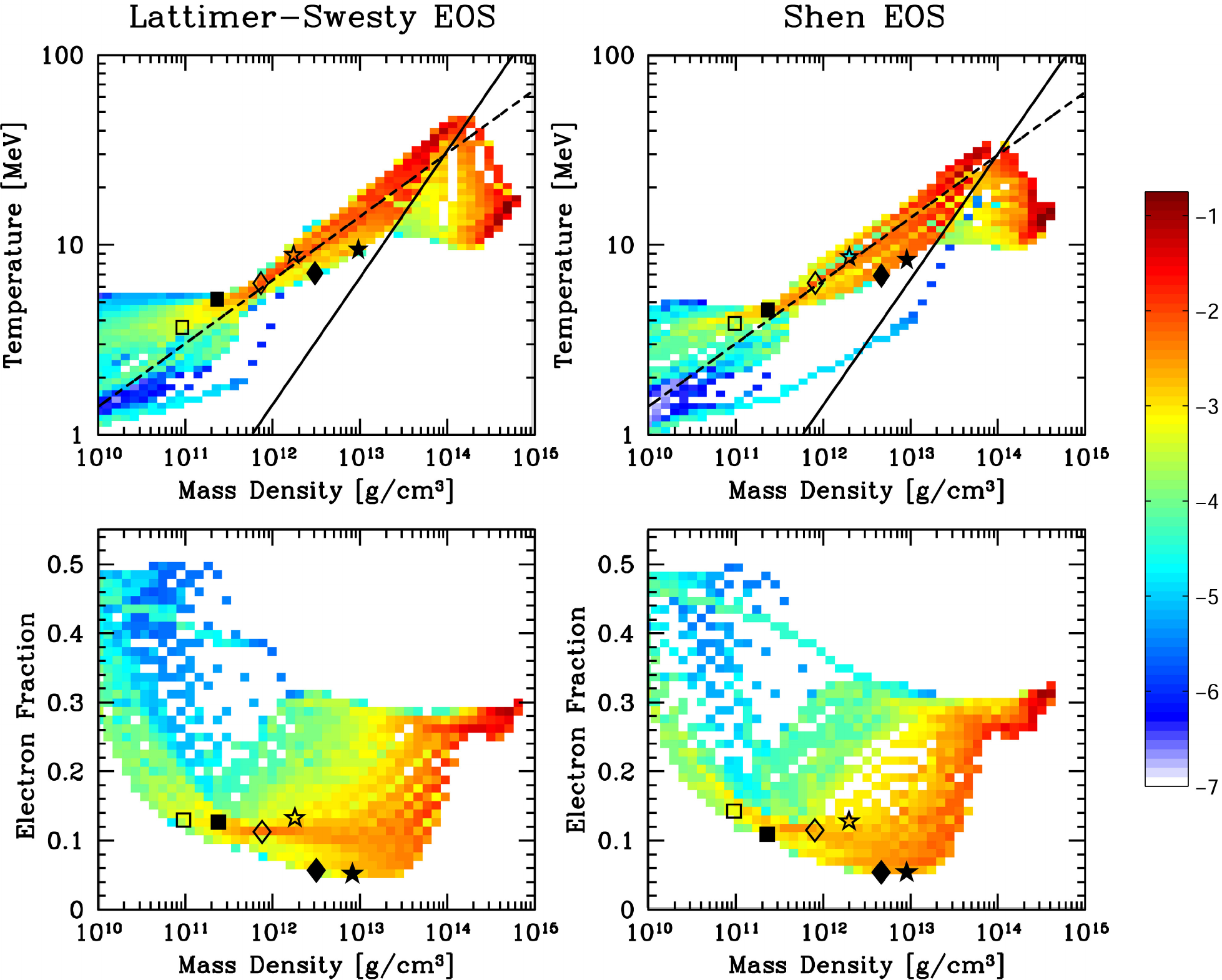}
\end{center}
\caption{Color-coded plots of the occurrence of thermodynamic
conditions during the post-bounce phase of spherically symmetric
supernova simulations launched from a $15 M_{\odot}$
progenitor star. Left (right) panels refer to a simulation with the
Lattimer-Swesty (Shen et al.) equation of state (EOS). Upper (lower) panels
show temperature $T$ (electron fraction $Y_e$) versus mass density
$\rho$. The occurrence is expressed on a logarithmic scale by the
amount of matter which experiences a specific set of conditions,
normalized to $M_{\odot}$, integrated over time during the
post-bounce phase [see text and Equation~(\ref{eq:histogram}) for
details]. The solid lines indicate where the temperature is equal to
the Fermi temperature of the baryons, $T_{\rm F} = (3\pi^2 \rho/m)^{2/3}/2m$,
and the dashed lines represent the conditions specified by 
Equation~(\ref{SNconditions}). In each
panel, six different symbols indicate the conditions under which
neutrinos decouple thermally from matter. Squares refer to $\nu_e$,
diamonds to $\bar{\nu}_e$, and stars to $\nu_{\mu,\tau}$ and
$\bar{\nu}_{\mu,\tau}$. Unfilled symbols refer to the early
post-bounce phase ($\approx 50 \, {\rm ms}$ after bounce), whereas
filled symbols refer to the latest times considered ($\approx 400 \,
{\rm ms}$ after bounce for the Lattimer-Swesty EOS simulation and
$\approx 430 \, {\rm ms}$ after bounce for the Shen et al.~case).}
\label{fig:conditions15}
\end{figure*}

\begin{figure*}[t]
\begin{center}
\includegraphics[trim=3mm 4mm 14mm 116mm,clip,scale=0.7,clip=]{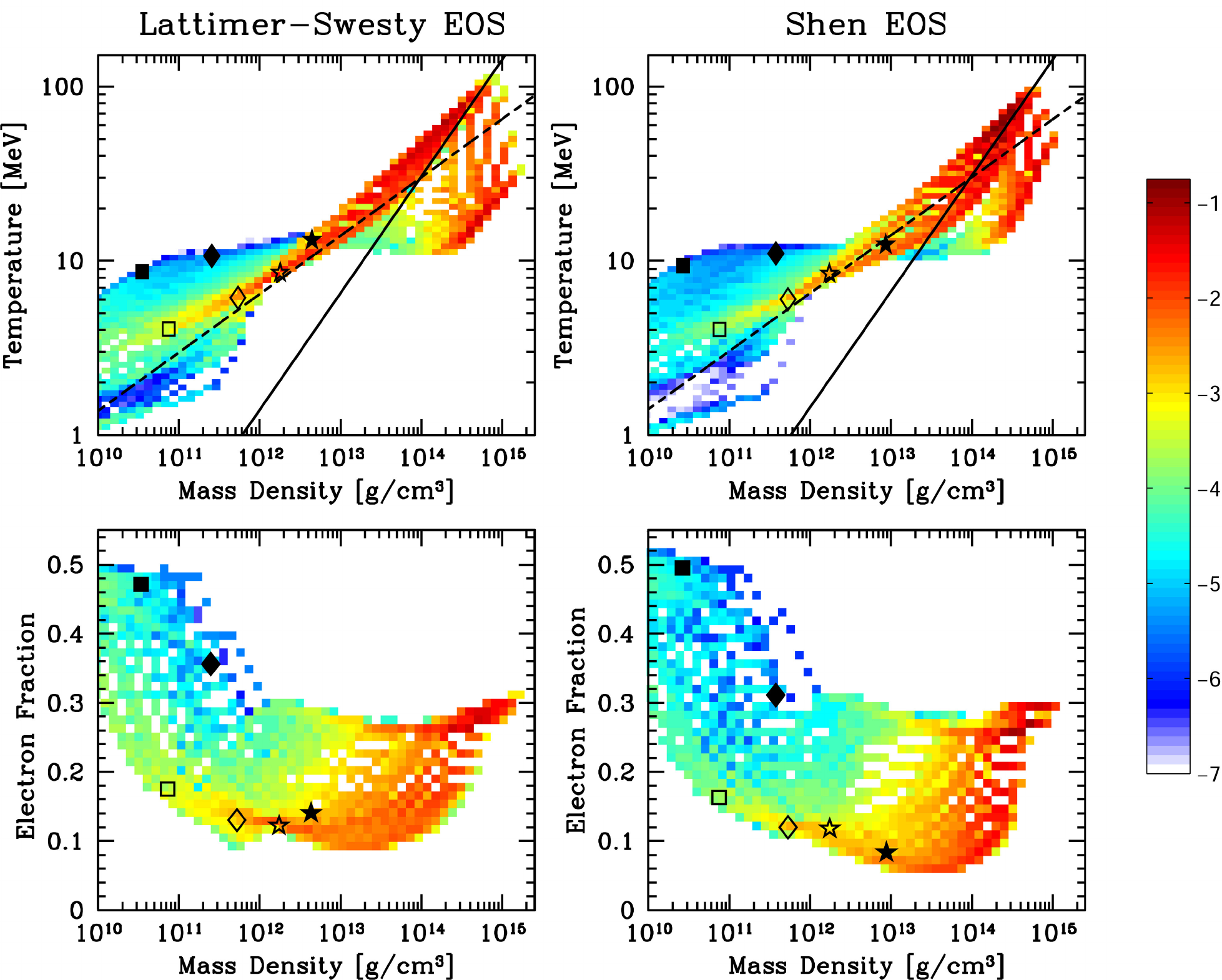}
\end{center}
\caption {Same as Figure~\ref{fig:conditions15}, but for a $40
M_{\odot}$ progenitor. As in Figure~\ref{fig:conditions15}, unfilled
symbols refer to the early post-bounce phase ($\approx 50 \, {\rm
ms}$ after bounce), whereas filled symbols refer to the latest times
considered, which is close to black-hole formation ($\approx 430 \,
{\rm ms}$ after bounce for the Lattimer-Swesty EOS simulation and
$\approx 1.36 \, {\rm s}$ after bounce for the Shen et al.~case).}
\label{fig:conditions40}
\end{figure*}

We consider the results of numerical simulations of the collapse of
$15 M_\odot$ and $40 M_\odot$ stars in a spherically symmetric
supernova model with Boltzmann neutrino transport
\citep{Fischer.Whitehouse.ea:2009}. Figure~\ref{fig:conditions15}
shows an overview of the conditions that are reached after the
collapse of the stellar core of a $15 M_{\odot}$ progenitor
star. The upper panels are constructed from a two-dimensional
histogram in the parameter space spanned by the matter density $\rho$
and the matter temperature $T$. The color of each bin
$[\rho_i,\rho_{i+1}]$, $[T_j,T_{j+1}]$ is chosen according to the
value of a time- and mass-weighted measure of the occurrence of the
conditions corresponding to the bin, where the labels on the color bar
specify the value of
\be
H(\rho,T) = \log_{10} \biggl[ \frac{1}{\tau M_{\odot}}
\int_{0}^{\tau}dt \int_{0}^{M} dM_{\rm encl} \: 
\chi_{ij}(t,M_{\rm encl}) \biggr] \,, 
\label{eq:histogram}
\ee
where $\chi_{ij}(t,M_{\rm encl}) = 1$ if the density and temperature
are reached in the bin, $\rho_{i} \leqslant \rho(t,M_{\rm encl}) <
\rho_{i+1}$ and $T_{j} \leqslant T(t,M_{\rm encl}) < T_{j+1}$, and
$\chi_{ij}(t,M_{\rm encl}) = 0$ otherwise. The integral over time $t$
runs from zero (at the time of core bounce) to the final time $\tau$,
while the integration over the enclosed mass $M_{\rm encl}(R) =
\int_0^R dr \, 4\pi r^2 \rho(r)$ runs from zero at
the center of the star to the total mass $M$ considered in the
computational domain. Hence, a white bin in the $\rho$--$T$-plane
indicates that the corresponding conditions rarely occur in the
simulation, while a red bin indicates conditions occurring for a large
mass and/or for a long time duration. The lower panels are constructed
in the same way, but with the temperature in
Equation~(\ref{eq:histogram}) replaced by the electron fraction
$Y_{e}$.  The symbols indicate where neutrinos decouple from matter.
This was determined by comparing the mean neutrino energy in the
simulation with the mean neutrino energy expected for neutrinos in
equilibrium at the temperature of the matter. The symbols mark the
conditions under which the difference between the two mean energies
was equal to $5\%$.

Figure~\ref{fig:conditions40} shows the conditions for a simulation
that was launched from a more massive progenitor star with a main
sequence mass of $40 M_{\odot}$. All panels show on the far right-hand
side the cold core of the protoneutron star at densities around $3
\times 10^{14} \gcmiq$, temperatures around $10-20 \mev$ and an
electron fraction just below $Y_{e}=0.3$. The black solid line shows
the baryon Fermi temperature as a function of density (see caption)
and thus the neutrons in this regime are degenerate. This matter lies
in the inner part of the stellar core, which is collapsing
subsonically, and it is unaffected by the bounce shock that forms
outside the core at the sonic point and runs outward. The compactness
of this highest density regime is sensitive to the equation of state
(EOS) of neutron-rich matter. In parametric EOSs, this depends mainly
on the incompressibility and the symmetry energy. If the EOS is stiff,
the mantle of the protoneutron star will lie at a larger radius in a
lower gravitational potential. This leads to comparatively small
neutrino luminosities and soft spectra. If the EOS is soft, the core
of the protoneutron star will be more compact and the mantle will lie
at a smaller radius deep in the gravitational potential. This leads to
larger neutrino luminosities and harder spectra. This general
relationship between neutrino emission and proto-neutron
star size applies to all factors that influence the compactness of
the core: general relativistic effects
\citep{Liebendoerfer.Mezzacappa.ea:2001,Bruenn.DeNisco.Mezzacappa:2001},
different progenitor masses \citep{Liebendoerfer.Mezzacappa.ea:2003},
and different EOSs \citep{Sumiyoshi.Yamada.Suzuki:2007}. In the left
panels of Figures~\ref{fig:conditions15} and~\ref{fig:conditions40}, we
show results from simulations based on the Lattimer-Swesty
EOS \citep{Lattimer.Swesty:1991}, which produces a more compact
protoneutron star core, while the right panels show results
from a simulation with the stiffer \citet{Shen.Toki.ea:1998b}~EOS.
New constraints from neutron matter calculations \citep{Kai}
and from modeling X-ray burst sources \citep{Steiner} favor EOSs
that yield more compact (cold) neutron stars than the
\citet{Shen.Toki.ea:1998b} EOS can give.

When the bounce shock runs through the mantle of the protoneutron
star, it causes a temperature jump. Hence, in the upper panels of
Figures~\ref{fig:conditions15} and~\ref{fig:conditions40}, matter in
the cold core is well separated from that in the shock-heated mantle,
which is found above the black line that indicates partially
degenerate conditions. The conditions in the mantle are represented by
a rather narrow band in the $\rho$--$T$-plane that runs from densities
around $10^{14} \gcmiq$ at temperatures of $\sim 30 \mev$ down to
densities of $10^{11} \gcmiq$ at temperatures of $\sim 3 \mev$.
This band is approximately described by the relation
\be
T = 3 \, \biggl( \frac{\rho}{10^{11} \gcmiq} \biggr)^{1/3}{\rm MeV} \,,
\label{SNconditions}
\ee
for which the temperature at $\rho =10^{14} \gcmiq$ is equal to the
Fermi temperature at that density. In Figures~\ref{fig:conditions15}
and~\ref{fig:conditions40}, this relation is shown as dashed lines. This
behavior may be understood as being a consequence of the fact that the
shocked matter expands approximately adiabatically (neglecting
neutrino heating). Due to the high temperature, most of the entropy
resides in the leptons and the photons, which behave as ideal,
relativistic gases, for which the entropy per unit volume varies as
$T^3$. Since the baryon density is proportional to $\rho$, the
temperature on adiabats behaves approximately as $T \sim \rho^{1/3}$.
Note that the maximum temperature
depends strongly on the mass of the progenitor star and, in the case
of the $15 M_{\odot}$ model, also on the EOS. During the first second
of the post-bounce phase, more than $70 \%$ of the $\nu_e$,
$\bar{\nu}_e$ and $\nu_{\mu,\tau}$
luminosities are emitted from this hot mantle of the protoneutron
star. At higher densities, the influence of the neutrinos on the
supernova dynamics is comparatively small, because neutrinos are
trapped so that thermal and weak equilibrium with the matter is
established. At very low densities, the influence of neutrino
processes is limited by small weak-interaction rates. The sizable
electron chemical potential in the mantle favors electron capture over
positron capture, so that it is the neutrino emitting protoneutron
star mantle that deleptonizes most efficiently and becomes the most
neutron-rich region in the computational domain. The resulting low
electron fractions in the density interval $\rho=10^{11} \gcmiq$ to
$10^{14} \gcmiq$ can be seen in the lower panels of
Figures~\ref{fig:conditions15} and~\ref{fig:conditions40}.

Neutrino reaction rates exhibit a strong energy dependence; as a
consequence, neutrinos with different energies decouple from the
matter at different locations. As an approximate measure of where
decoupling occurs, we take the radius at which the mean neutrino
energy obtained from the simulation begins to differ from the one
expected for a gas of neutrinos in thermal equilibrium with the
matter. The symbols in Figures~\ref{fig:conditions15}
and~\ref{fig:conditions40} indicate the thermodynamic conditions at
these points. To indicate the time evolution, the conditions for
decoupling are shown at $50 \, {\rm ms}$ after bounce (open symbols)
and at the end of the simulation (filled symbols). In general, the
neutrinospheres for electron neutrinos (square symbols) are found at
the lowest density due to their charged-current reactions with
neutrons, which are the most abundant species under the given
conditions. Electron antineutrinos, which have charged-current
reactions with protons in addition to neutral-current reactions,
decouple on average at densities ten times larger (diamonds). The
$\mu$ and $\tau$ neutrinos and antineutrinos interact with matter only
via neutral-current reactions and therefore decouple at the highest
density, as indicated by the stars. This density hierarchy translates
into a temperature hierarchy for the regions from which neutrinos are
emitted. As a result, the spectra of $\nu_{\mu,\tau}$ and
$\bar{\nu}_{\mu,\tau}$ are harder than that of $\bar{\nu}_e$, which in
turn is harder than that of $\nu_e$.

The post-bounce evolution can be separated into two phases. During the
first phase the mass of the protoneutron star is significantly lower
than the maximum mass supported by the EOS. Thus, the protoneutron
star does not contract significantly and the neutrinospheres move to
higher density with time as the accretion rate and the density of the
accreting layers decrease. Because regions at higher density have
higher temperature, this also leads to a continuous increase of the
temperature of emitted neutrinos. However, in spherically symmetric
models it can happen that the steady cooling by $\nu_{\mu,\tau}$ and
$\bar{\nu}_{\mu,\tau}$ emission within a narrow range of the mass
coordinate decreases the local temperature sufficiently that the
mean energy of $\nu_{\mu,\tau}$ and $\bar{\nu}_{\mu,\tau}$ decreases
and eventually falls below that of the $\bar{\nu}_e$
\citep{Liebendoerfer.Messer.ea:2004}. The first phase of the
post-bounce evolution is well represented by the $15 M_{\odot}$ model
shown in Figure~\ref{fig:conditions15}. After the onset of the supernova
explosion, which was not successful for the models shown in
Figures~\ref{fig:conditions15} and~\ref{fig:conditions40}, the mean
energies of the neutrino spectra decrease as the protoneutron star
cools \citep{Fischer.Whitehouse.ea:2009}. Due to the lower opacity for
neutrinos with lower energy, the neutrinospheres continue to
shift to higher density and make the high-density part of the
protoneutron star more `neutrino-visible'.

The first phase of the postbounce evolution dictates the success of
the supernova explosion, i.e., whether sufficient entropy can be
accumulated behind the standing accretion shock to revive it and drive
it through the outer layers of the star. This phase of the supernova
mechanism is extremely challenging to model because the emission,
transport and absorption of neutrinos behind the shock couples to
vigorous three-dimensional fluid instabilities between the
neutrinospheres and the accretion shock. These multidimensional
effects are not included in the data shown in
Figures~\ref{fig:conditions15} and~\ref{fig:conditions40}.

If an explosion has not taken place during the first stage, the
post-bounce evolution enters a second phase when the mass of the
protoneutron star approaches the maximum stable mass. This scenario
may occur in the rarer cases of `failed supernovae' that are
triggered by the collapse of a very massive progenitor star. As the
limiting stable mass of the protoneutron star is approached, accretion
makes the protoneutron star more compact and pushes matter with high
entropy into the regime of the neutrinospheres. The corresponding
temperature change increases the opacity so that the neutrinospheres
shift to lower densities, as can be seen in
Figure~\ref{fig:conditions40}. This leads to a dramatic increase in
the mean energy of emitted $\nu_{\mu,\tau}$ and $\bar{\nu}_{\mu,\tau}$
\citep{Liebendoerfer.Messer.ea:2004,Sumiyoshi.Yamada.Suzuki:2007}. However,
only the $40 M_{\odot}$ model reaches this second phase within the
duration of the simulation. With the Lattimer-Swesty EOS, it takes
$0.43 \, {\rm s}$ to form a black hole, while for the stiffer Shen et
al.~EOS it takes $1.4 \, {\rm s}$
\citep{Sumiyoshi.Yamada.Suzuki:2007,Fischer.Whitehouse.ea:2009}.

In summary, the emission of neutrinos from the partially degenerate
hot mantle of the protoneutron star, the EOS at nuclear densities, and
the accretion rate set by the progress of the supernova shock and the
density of the outer layers of the
progenitor star are the key ingredients that determine the
compactification of the protoneutron star during the early post-bounce
evolution. The emission of neutrinos also shapes the radial entropy
and lepton number profiles that determine where stratified layers of
matter are stable against convection. The emitted neutrinos finally
can be absorbed by matter at larger radii between the protoneutron
star mantle and the standing accretion shock. The heat transferred per
absorbed neutrino increases with the difference between the
temperature of the neutrinos and that of the absorbing matter.  This
difference increases with the distance from the neutrinospheres, but
the neutrino number flux decreases with the distance because of
geometrical dilution. Hence, there is a region close to the
protoneutron star where heating is most effective. This peak heating
at the base of the matter lying between the protoneutron star and the
standing accretion shock establishes a negative entropy gradient. The
entropy gradient induces fluid instabilities that start to transport
energy to the shock by mechanical turnover
\citep{Herant.Benz.ea:1994,Marek.Janka:2009}. The perturbations in the
convective fluid flow are enhanced by the standing accretion shock
instability
\citep{Blondin.Mezzacappa.DeMarino:2003,Foglizzo.Galletti.ea:2007}.
The fluid instabilities and the net rate of neutrino cooling and
heating determine the further evolution and geometry of the
supernova. It is therefore essential for supernova models to
accurately treat the neutrino-matter interactions in the hot mantle of
the protoneutron star under the conditions shown in
Figures~\ref{fig:conditions15} and~\ref{fig:conditions40}, which we
shall investigate in Section~\ref{results}.

\section{General formalism}
\label{formalism} 

Motivated by the relevant supernova conditions, we extend our work on
neutrino processes in degenerate matter \citep{LPS,Bacca} to the case
of partially degenerate and nondegenerate neutrons.

The basic ingredients in the calculation of rates of neutrino
processes are the structure factors for the vector and axial vector
response \citep{Raffelt}. For nonrelativistic neutrons, the vector
current couples to the neutron density and the axial current to the
neutron spin. We denote the coupling strengths by $C_V$ and $C_A$,
respectively, and for free neutrons $C_V = −1/2$ and $C_A = − g_A/2 =
−1.26/2$. We shall focus on the axial current response for two
reasons. The first is that it is responsible for the largest
contributions to rates of neutrino processes. The second is that axial
current processes have potential for equilibrating neutrino energy
distributions because the axial current is not conserved and therefore
it is possible for the energy transfer to or from the neutrinos to be
nonzero even for processes in which there is a small momentum
transfer. For the vector interaction, the energy transfer vanishes for
zero momentum transfer because the vector current is conserved. For
neutrons, the structure factor for the axial current therefore has the
form
\be
S_{\rm A}(\om,\qq) = C_A^2 \, S_\sigma(\om,\qq) \,,
\ee
where the spin structure factor is given by
\be
S_{\sigma}(\om,\qq) = \frac{1}{\pi n} \, \frac{1}{1-\rme^{-\om/T}} \,
{\rm Im} \, \chi_\sigma(\om,{\bf q}) \,.
\label{structspin}
\ee
Here $\chi_\sigma$ is the spin-density--spin-density response 
function. (We use units in which $\hbar=c=k_{\rm B} = 1$.)

We begin by presenting a phenomenological approach. As has been
stressed by Raffelt and coworkers, it is a good first approximation in
many situations to consider the long-wavelength response because the
momentum transfer to the nucleons is of the order of a typical neutrino
momentum, which is generally small compared with the characteristic
momentum of a nucleon. The latter is of the order of the Fermi momentum $(3\pi^2
n)^{1/3}$ for degenerate matter and the thermal momentum $(mT)^{1/2}$,
where $m$ is the nucleon mass, for nondegenerate matter. We consider
the relaxation of $S$, the total $z$-component of the spin of the
system to its equilibrium value $S_{\rm eq}$, which is given in linear
response theory by
\be
S_{\rm eq} = -\chi_M \, U_z \,, 
\label{eqmS}
\ee  
where $-U_z$ corresponds to the strength of a magnetic field that
polarizes the system and $\chi_M$ is, apart from factors, the
static magnetic susceptibility.  If one assumes that the approach
of $S$ to its equilibrium value is proportional to the difference
between $S$ and its equilibrium value, we may write 
\be
\frac{dS}{dt} = -\frac{S-S_{\rm eq}}{\tau} \,,
\label{relax}
\ee 
where $\tau$ is a characteristic relaxation time. The quantity
$1/\tau$ corresponds to what \citet{HR} refer to as the
spin fluctuation rate. The total spin of
the system thus approaches its equilibrium value exponentially, if 
the magnetic field does not depend on time.

On Fourier transforming Equations~(\ref{eqmS}) and~(\ref{relax})
one finds for the frequency-dependent response function  
\be
\chi_\sigma(\omega) \equiv -\frac{S}{U_z} = 
\frac{\chi_M}{1-\rmi \omega \tau} \,.
\label{susc}
\ee
In order to make detailed calculations, we shall consider nucleonic
matter as a system of interacting quasiparticles. We shall assume that
the spectrum of a single quasiparticle may be described by an
effective mass $m^*$ that is independent of its momentum $\pp$, and
therefore the energy of a single quasiparticle is given by
\be
\epsilon_{\pp \sigma} = \epsilon_0 + \frac{p^2}{2 m^*} \,.
\ee
We take into account two-body quasiparticle interactions, with a
spin-dependent part of the form 
\be
f_{\pp {\bm \sigma}_1,\pp' {\bm \sigma}_2} =g_0 \, {\bm \sigma}_1
\cdot {\bm \sigma}_2 \,,
\ee
where ${\bm \sigma}_i$ are Pauli matrices.
The theory may be regarded as an extension of Landau's theory of Fermi
liquids generalized to higher temperatures when the quasiparticle
interaction is independent of the momenta of the quasiparticles.  The
parameters $m^*$ and $g_0$  generally depend on temperature and
density, and their values are chosen to ensure that bulk
properties such as the specific heat and magnetic susceptibility agree
with the results of more microscopic calculations. 

We shall take the coupling of the spin to the external field $U_z$ to
be given by a change in the energy of a quasiparticle, $\epsilon_{\pp
\sigma}$, equal to 
\be
\delta \epsilon_{\pp \sigma} = \sigma \, U_z \,.
\ee
By generalizing the standard calculation of the static magnetic
susceptibility \citep{BaymPethick} to higher temperatures, one finds 
\be
\chi_M = \frac{\chi^0}{1+g_0 \chi^0} \,,
\ee
where
\be
\chi^0 = -\sum_{\pp\sigma} \frac{\partial 
n(\epsilon_{\pp\sigma})}{\partial \epsilon_{\pp\sigma}} \,,
\label{staticsusc}
\ee  
is the static susceptibility in the absence of interactions between
quasiparticles and 
\be
n(\epsilon) = \frac1{\rme^{(\epsilon-\mu)/T}+1} \,,
\ee
is the Fermi function, $\mu$ being the chemical potential. At
temperatures low compared with the Fermi degeneracy temperature
$T_{\rm F} = p_{\rm F}^2/2m^*$, where $p_{\rm F}$ is the Fermi momentum,
the density of states
$\chi^0 = m^* p_{\rm F}/\pi^2$ for a single species of particles with
two spin states, while at temperatures high compared with $T_{\rm F}$,
one finds the Curie law $\chi^0 = n/T$. On inserting
Equation~(\ref{staticsusc}) into Equation~(\ref{susc}) one finds
\be  
\chi_\sigma(\omega)=\frac{\chi^0}{1+g_0 \chi^0-\rmi \omega \tau_\sigma} \,,
\label{susc2}
\ee
where 
\be
\tau_\sigma = \tau \, (1+g_0 \chi^0) \,.
\ee
This equation has the same form as in \citet{LPS} [see, e.g.,
Equation~(24)]. The significance of the time $\tau_\sigma$ may be
brought out by writing Equation~(\ref{relax}) in terms of the
deviation of the spin from its local equilibrium (l.e.) value 
\be
S_{\rm l.e.} = \chi^0 (-U_z+g_0 \, S) \,,
\ee
which is the equilibrium spin density in a system exposed to an
external field $-U_z$ plus a ``molecular field'' $g_0 \, S$ due to
interactions with the other particles. Equation~(\ref{relax}) then
becomes 
\be
\frac{dS}{dt} = -\frac{S-S_{\rm l.e.}}{\tau_\sigma} \,.
\label{relaxle}
\ee
From kinetic theory, one saw in the degenerate limit how mean-field
effects enter relaxation rates \citep{LPS}: the time $\tau_\sigma$
does not depend on mean-field effects, while the characteristic time
for relaxation of a deviation of the total spin from its equilibrium
value is decreased by the factor by which mean-field effects decrease
the magnetic susceptibility.  The reason for this effect is that when
the magnetic susceptibility is reduced, the free energy difference
driving the return to equilibrium, which for a given spin deviation is
inversely proportional to the magnetic susceptibility, is increased, and
relaxation is faster.

In the work of \citet{HR}, the normalization of the spin response
function was determined by the condition that it give correctly the
static structure factor, which was assumed to be that of a
noninteracting gas.  It is therefore of interest to investigate the
relationship between the static structure factor and the spin response
function when interactions are taken into account. The static
structure factor $S_\sigma(\qq)$ is defined by
\be
S_\sigma(\qq) = \int _{-\infty}^\infty d\omega \, S_\sigma(\omega,\qq) \,.
\ee
Since the dynamical structure factor $S(\omega,\qq)$ is related to the
response function by Equation~(\ref{structspin}), one has
\be
S_\sigma(\qq) = \frac{1}{n} \int_{-\infty}^\infty \frac{d\omega}{\pi}
\, \frac{{\rm Im}\chi_\sigma(\omega,\qq)}{1-\rme^{-\omega/T}}
= \frac{1}{n} \int_{-\infty}^\infty \frac{d\omega}{2\pi} \,
{\rm Im}\chi_\sigma(\omega,\qq) \, \coth[\omega/(2T)] \,,
\ee
where we have used the fact that ${\rm Im}\chi_\sigma(\omega,\qq)$ is an odd
function of $\omega$. From the inequality $\bigl|\coth[\omega/(2T)]
\bigr| \geqslant |2T/\omega|$, it follows that
\be
S_\sigma(\qq) \geqslant \frac{T}{n} \int_{-\infty}^\infty
\frac{d\omega}{\pi} \, \frac{{\rm Im}\chi_\sigma(\omega,\qq)}{
\omega} = \frac{T}{n} \, \chi_M \,,
\label{inequality}
\ee
the equality applying only if all the weight of ${\rm Im}\chi_\sigma$
is at zero frequency. Using Equation~(\ref{susc2}) for the
response function, one finds
\be
S_\sigma(\qq \rightarrow 0) \geqslant \frac{T}{n} \, 
\frac{\chi^0}{1+g_0 \chi^0} \,.
\ee
These results represent a generalization of Equation~(15) of
\citet{HR} to allow for spin correlations. For neutron matter at low
temperature, $g_0\chi^0$ is equal to the Landau parameter $G_0$, which
is calculated to be approximately 0.45 at a density of $\rho_0/100$
and 0.8 at nuclear saturation density, $\rho_0 = 2.8 \times 10^{14}
\gcmiq$ \citep{SchwenkBF}, and therefore interactions
depress the static structure factor by up to almost a factor of
two. For density correlations, the equality sign applies in the
long-wavelength limit, since total particle number is conserved and,
consequently, matrix elements of the density operator to states with
nonzero excitation energy vanish in this limit
\citep[Section~1.3.3]{BaymPethick}. If interactions that do not
conserve spin are neglected, the inequality~(\ref{inequality})
becomes an equality at long wavelengths. This approximation has been
employed in calculations of neutrino processes in nondegenerate matter
at low densities based on the virial expansion \citep{horowitz}.

A number of cautionary remarks about our approach should be made.
First, in the above we have calculated only the contribution to the
susceptibility due to the coupling of the external field to single
quasiparticle-quasihole intermediate states. The effects of multipair
states are taken into account in so far as they may be regarded as
arising from single-pair states as a consequence of the collision term
in the kinetic equation. Other processes that can result in the creation
of two- and higher-quasiparticle-quasihole states, e.g.,
two-body contributions to effective operators \citep{Javier}, have
been neglected. As emphasized in \citet{OlssonHaenselPethick}, because
of the noncentral character of nuclear forces, these are nonzero even
for very long wavelengths of the exciting field. Similarly, we have
neglected many-body contributions that can renormalize the magnetic
moment (or in the case of the coupling to the weak field, the weak
charge) of the nucleon. In estimating rates of processes one must
multiply the above results by the square of the appropriate
renormalization factors.

One may ask how good is the assumption of a single relaxation
time. This question was addressed for degenerate nucleons in
\citet{PS}.  There it was shown for $\omega\ll T$ that, if the
relaxation time is chosen to reproduce the leading behavior for
$\omega\gg 1/\tau_\sigma$, the real and imaginary parts of the
response function given by Equation\ (\ref{susc2}) differ from the
result obtained by solving the transport equation exactly by less than
10\%.  For nondegenerate matter this question has not yet been
addressed, but one can draw on experience with other transport
problems for this case.  The relaxation time at high values of
$\omega\tau_\sigma$ is given by the simplest variational estimate for
the relaxation time, which uses a trial function proportional to
the quasiparticle spin.
The actual relaxation time at low frequencies is longer
than this estimate, but to date no estimates of it have been made.
For nondegenerate, spinless atomic gases one finds that relaxation
times for viscosity and thermal conduction in the hydrodynamic regime
differ from the simplest variational estimates by amounts that are
typically of order two per cent or less \citep{chapman}.  The part of
the interaction relevant for spin relaxation depends on the spin of
the particles, and it has a different spatial dependence from typical
interactions between spinless atoms, so it is desirable to calculate
the ratio of the relaxation times in the hydrodynamic and
collisionless regimes for nuclear interactions.  However, it seems
unlikely that the deviations will be more than a few per cent, which
is small compared with the uncertainties in the strong scattering amplitudes
that enter the relaxation rates. For practical purposes it is
therefore expected to be a very good approximation to use
Equation~(\ref{susc2}) for all $\omega \tau_\sigma$.

\section{Relaxation time}
\label{time}

In this section, we calculate the response function microscopically
for frequencies high compared to the collision rate.\footnote{Here we
are referring only to the quasiparticle contribution to the
response. By large frequencies, we mean frequencies large compared
with $1/\tau_\sigma$, but still small compared to the other energy
scales in the system.}
In terms of the phenomenological model discussed in the previous 
section, this is given by
\be
\chi(\omega) \simeq \rmi \, \frac{\chi^0}{\omega \tau_\sigma} \,.
\label{suscHigh}
\ee

\begin{figure*}[t]
\begin{center}
\includegraphics[scale=1.1,clip=]{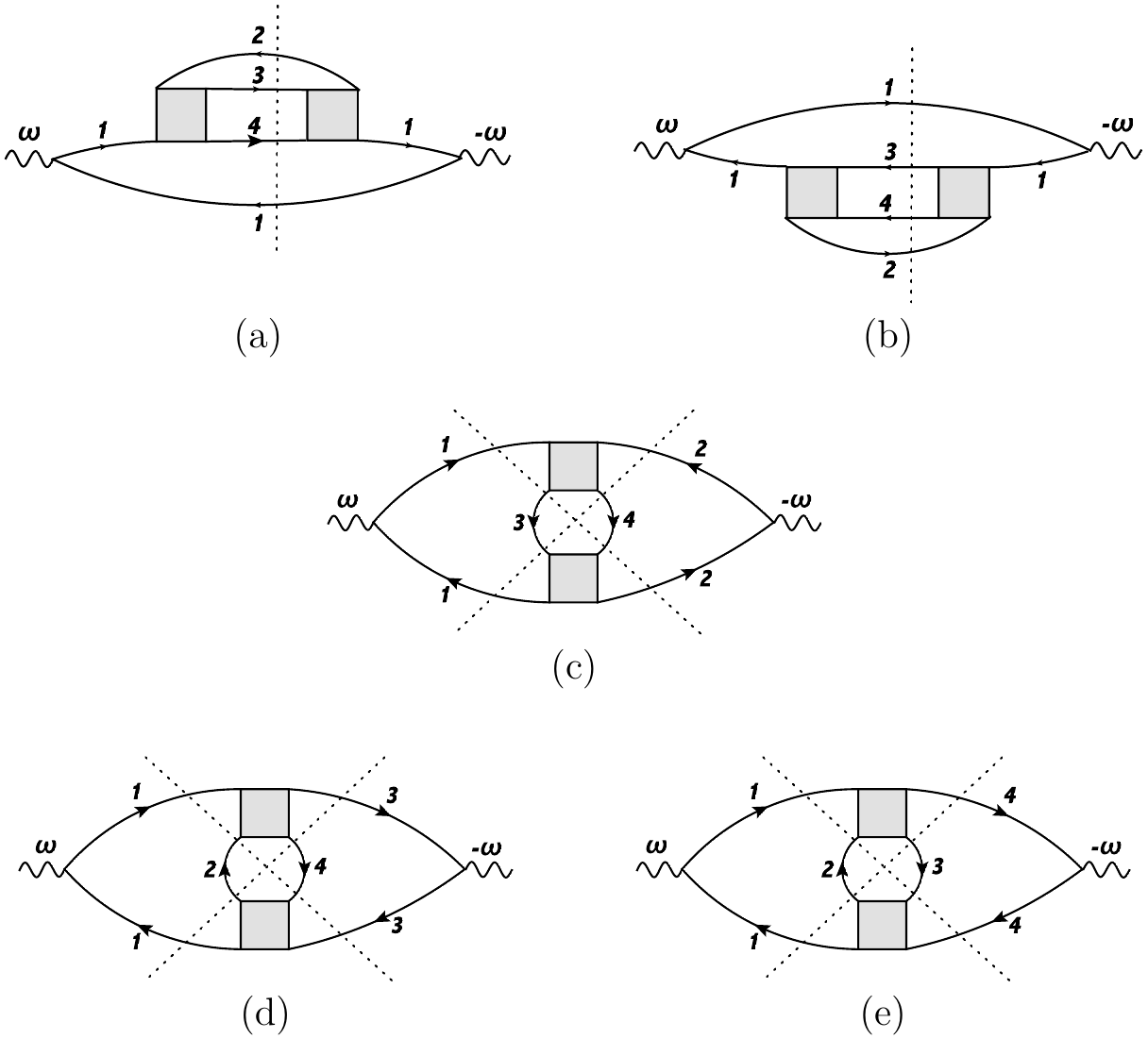}
\end{center}
\caption{Diagrams for the imaginary part of the response function.
Solid lines represent quasiparticle propagators and the square
boxes interactions of quasiparticles. The dashed lines represent
two-quasiparticle--two-quasihole states that are on the energy
shell.\label{diagrams}}
\end{figure*}

Quantitatively, the response function may be calculated using
field-theoretical methods, as described in \citet{LOP}.  The
contribution to the response function proportional to $1/\omega$
follows from the diagrams shown in Figure~\ref{diagrams}. In the figure,
the lines represent quasiparticles, and the interaction vertices are
those for quasiparticles.  The square boxes denote scattering vertices
for quasiparticles, symmetrized with respect to the incoming and
outgoing states. The diagrams in Figure~\ref{diagrams}~(a) correspond to
self-energy corrections which take into account the nonzero width of
the quasiparticles, while those in Figure~\ref{diagrams}~(b)--(d)
correspond to vertex corrections.

As we have argued above, the leading contribution to the response
functions at large $\omega$ is imaginary, and a convenient way of
calculating this is by use of the techniques developed in field theory
\citep{landaucuts,cutkosky,taylor}, which have been exploited in the
context of condensed matter in \citet{langer1,langer2} and
\citet{carneiro}. The energy denominators that lead to singularities
may be identified by finding intermediate states in the response
function which have the property that when lines corresponding to the
intermediate state are cut, the diagram falls into just two parts.  In
Figure~\ref{diagrams} such intermediate states corresponding to
two-quasiparticle--two-quasihole states are denoted by dashed
lines. By including all such intermediate states one automatically
includes contributions that correspond both to ``in-scattering'' and
``out-scattering'' terms in an approach based on a kinetic
equation. On evaluating the response function one finds
\begin{align}
\chi(\omega) &= \rmi \, \frac{2\pi}{\omega^2} \, \frac{1}{2} \, 
\sum_{1234} \sigma_1 (\sigma_1 + \sigma_2 - \sigma_3 -\sigma_4) \nonumber \\
&\quad \times \Bigl[ \bigl| \langle 34 |{\cal T}| 12 \rangle \bigr|^2 
n_1 n_2 (1-n_3)(1-n_4)
-\bigl| \langle 12 |{\cal T}| 34 \rangle \bigr|^2
(1-n_1)(1-n_2) n_3 n_4 ) \Bigr] 
\delta^{+}_\epsilon \delta_\pp \,.
\label{response}
\end{align}
Here the index 1 is shorthand for $\pp_1\sigma_1$ and so on, $n_i$ are
equilibrium distribution functions, $\delta_\pp \equiv \delta ( \pp_1
+\pp_2-\pp_3-\pp_4)$ and $\delta^{\pm}_\epsilon \equiv
\delta(\pm\omega + \epsilon_1+\epsilon_2-\epsilon_3-\epsilon_4)$, and
we have dropped the momentum transfer $\qq$ from the momentum
conservation condition according to the arguments given above.  The
factor of $1/2$ is to avoid double counting of final states when the
sums over $\pp_3$ and $\pp_4$ are unrestricted. The two different
scattering matrices in Equation~(\ref{response}) may be combined
if one makes use of the fact that the interaction is invariant under
time reversal, which implies
\be
\langle 34 |{\cal T}| 12 \rangle = \langle -1,-2 |{\cal T}| -3,-4 
\rangle^* \,,
\ee
where the minus sign indicates that both the momentum of a
quasiparticle and its spin must be reversed. Since the distribution
function depends on momentum only through its magnitude, since all
momenta are summed over, and since each term is quadratic in the spins
of quasiparticles, it follows that $\chi(\omega)$ has the form
of Equation~(\ref{suscHigh}) with
\be
\frac{1}{\tau_\sigma} = \frac{\pi}{\chi^0 \, \omega} \,
\sum_{1234} \sigma_1(\sigma_1-\sigma_2-\sigma_3-\sigma_4)
\bigl| \langle 34 |{\cal T}| 12 \rangle \bigr|^2 \,
n_1 n_2 (1-n_3)(1-n_4) \, \delta_\pp \,
\bigl[\delta^+_\epsilon - \delta^-_\epsilon\bigr] \,.
\label{tausigmafinal}
\ee
This form demonstrates explicitly the fact that the relaxation time is
an even function of $\omega$, as is required by the condition that the
real part of $\chi$ be an even function of $\omega$ and the imaginary
part an odd function. By making use of the relation
\be
\frac{n_i}{1-n_i} = \rme^{-(\epsilon_i-\mu)/T} \,,
\ee
and exchanging the roles of 1 and 2 with 3 and 4 in the sum in
Equation~(\ref{tausigmafinal}), one can see that the second term is
$\rme^{-\omega/T}$ times the first one, and therefore
\be
\frac{1}{\tau_\sigma} = \frac{(1-\rme^{-\omega/T})}{\omega} \,
\frac{\pi}{\chi^0} \, \sum_{1234} \sigma_1
(\sigma_1+\sigma_2-\sigma_3-\sigma_4)
\bigl| \langle 34 |{\cal T}| 12 \rangle \bigr|^2 \, 
n_1 n_2 (1-n_3)(1-n_4) \, \delta_\pp \, \delta^+_\epsilon \,.  
\ee
This may be expressed in the more symmetrical form 
\be
\frac{1}{\tau_\sigma} = \frac{2 \sinh[\omega/(2T)]}{\omega} \,
\frac{\pi}{\chi^0} \, \sum_{1234} \sigma_1
(\sigma_1+\sigma_2-\sigma_3-\sigma_4)
\bigl| \langle 34 |{\cal T}| 12 \rangle \bigr|^2 \, 
\delta_\pp \, \delta^+_\epsilon
\prod_{i=1}^4 \, \frac{1}{2\cosh[(\epsilon_i-\mu)/(2T)]} \,.  
\ee
To convert the sums over momenta to integrals, we use as momentum
variables the total momentum $\PP=\pp_1+\pp_2=\pp_3+\pp_4$ and the
relative momenta $\pp=(\pp_1-\pp_2)/2$ and $\pp'=(\pp_3-\pp_4)/2$.
Because all directions of the momenta are being integrated over, it is
convenient to work with the quantity
\be
W = \frac{1}{12} \, \sum\limits_{k=1,2,3}
{\rm Tr} \Bigl[ \, \langle 34 |{\cal T}| 12 \rangle^* \,
{\bm \sigma}_1^k \, \bigl[ ({\bm \sigma}_1 + {\bm \sigma}_2)^k \, , \,
\langle 34 |{\cal T}| 12 \rangle \bigr] \, \Bigr] \,,
\ee 
which depends only on $P, p$, and $p'$, and three angles specifying the
relative orientation of the vectors $\PP, \pp$, and $\pp'$. The trace
${\rm Tr}$ is over the spin indices for all the quasiparticles and
the spin structure with the commutator is the same as for degenerate
conditions \citep{LPS}. The integration over $p'$ may be eliminated by
making use of the energy conservation condition,
\be
\frac{p'^2}{m^*} = \frac{p^2}{m^*} + \omega \,.
\ee
The final result is
\begin{align}
\frac{1}{\tau_\sigma} &= \frac{1}{2^5\pi^6}
\frac{2\sinh[\omega/(2T)]}{\omega} \, \frac{m^*}{\chi^0}
\int_{-1}^1 d\cos\theta \int_{-1}^1 d\cos\theta' \int_0^{2\pi} d\phi
\nonumber \\
&\quad \times \int_0^\infty dP P^2 \int_0^\infty dp \, p^2 \,
(p^2+m^*\omega)^{1/2} \, W \, \prod_{i=1}^4 \, \frac{1}{2
\cosh[(\epsilon_i-\mu)/(2T)]} \,,
\label{tausigmafinal2}
\end{align}
where $\theta$ is the angle between $\PP$ and $\pp$, $\theta'$ the
angle between $\PP$ and $\pp'$, and $\phi$ is the angle between the
normal to the plane containing $\PP$ and $\pp$ and the normal to the
plane containing $\PP$ and $\pp'$. For definiteness, we have taken
$\omega$ to be positive. For negative $\omega$, the lower limit of the
integral over $p$ is $(m^*|\omega|)^{1/2}$. The integral is
symmetrical in $p$ and $p'$ since $2dp \, p^2 \,
(p^2+m^*\omega)^{1/2} = pp'dp^2 = pp'dp'^2$.
 
Alternative forms for the relaxation rate that bring out more clearly
the physical origin of the various contributions are obtained by
replacing the factors
\be
2\sinh[\omega/(2T)] \, (p^2+m^*\omega)^{1/2} \, W \, \prod_{i=1}^4 \,
\frac{1}{2 \cosh[(\epsilon_i-\mu)/(2T)]} \,,
\ee
in Equation~(\ref{tausigmafinal2}) by
\be
(p^2+m^*\omega)^{1/2} \, W \,
\bigl[ \, n_1n_2(1-n_3)(1-n_4)-(1-n_1)(1-n_2)n_3 n_4 \, \bigr] \,,
\ee
or by
\be
\bigl[(p^2+m^*\omega)^{1/2}-(p^2-m^*\omega)^{1/2}\bigr] \, W \,
n_1 n_2(1-n_3)(1-n_4) \,.
\ee

\subsection{Nondegenerate limit}
\label{nondegen} 
 
In the nondegenerate limit, $\mu/T\rightarrow -\infty$, the product of
thermal factors in Equation~(\ref{tausigmafinal2}) becomes
\be
\rme^{-(\epsilon_1+\epsilon_2+\epsilon_3+\epsilon_4)/(2T)} \,
\rme^{2\mu/T} = \rme^{-[P^2/(4m^*)+p^2/m^*+\omega/2]/T} \,
\rme^{2(\mu-\epsilon_0)/T} \,,
\ee
which shows that the distribution functions for the relative momentum
and the center-of-mass momentum are independent of each other.
However, the integral in Equation~(\ref{tausigmafinal2}) is not
particularly simple, because $W$ is generally a function of all 5
integration variables. Simplification is possible if the transition
probability $W$ is independent of the center-of-mass momentum, which is the
case if the influence of the medium on the scattering process is
neglected. Then $W$ is a function of $p$, $p'=(p^2+m^*\omega)^{1/2}$,
and $\Theta$, the angle between $\pp$ and $\pp'$. The expression for
the relaxation rate then takes the form
\begin{align}
\frac1{\tau_\sigma} &= \frac{1}{8\pi^5} \,
\frac{2\sinh[\omega/(2T)]}{\omega} \, \frac{m^*}{\chi^0} \, 
\rme^{2(\mu-\epsilon_0)/T} \int_{-1}^1 d\cos\Theta \int_0^\infty dP
P^2 \, \rme^{-P^2/(4m^*T)} \nonumber \\[1mm]
&\quad \times \int_0^\infty  dp \, p^2 \, \rme^{-p^2/(m^*T)} \, 
(p^2+m^*\omega)^{1/2} \, W \, \rme^{-\omega/(2T)} \,, \\[2mm]
&= \frac{m^* n}{4\pi} \, \frac{2\sinh[\omega/(2T)]}{\omega/T} \,
\langle (p^2+m^*\omega)^{1/2} \, W \, \rme^{-\omega/(2T)} \rangle \,,
\label{tausigmafinalnondegen}
\end{align}
where
\begin{align}
\langle \ldots \rangle &= 
\cfrac{\int_0^\infty dp \, p^2 \, \rme^{-p^2/(m^*T)} \int_{-1}^1 d\cos\Theta 
\, \ldots}{\int_0^\infty dp \, p^2 \, \rme^{-p^2/(m^*T)} \int_{-1}^1 d\cos
\Theta} \,, \\[1mm]
&= \frac{2}{\sqrt{\pi} (m^* T)^{3/2}} \int_0^\infty dp \, p^2 \, 
\rme^{-p^2/(m^*T)} \int_{-1}^1 d\cos\Theta \, \ldots \,.
\end{align}
In the nondegenerate limit, the dependence on density factors from the
integral. We therefore introduce the quantity
\be
\Xi(\omega) \equiv \frac{2\sinh[\omega/(2T)]}{\omega/T} \,
\langle (p^2+m^*\omega)^{1/2} \, W \, \rme^{-\omega/(2T)} \rangle \,,
\label{defX}
\ee
in terms of which the spin relaxation rate is given by
\be
\frac{1}{\tau_\sigma} = \frac{m^* n}{4\pi} \: \Xi(\omega) = 
\frac{\rho \, m^*/m}{4\pi} \: \Xi(\omega) \,.
\label{tauX}
\ee

\section{Results}
\label{results}

We next calculate the spin relaxation rates using nucleon-nucleon (NN)
interactions based on chiral EFT to
next-to-next-to-next-to-leading order (N$^3$LO), as in \citet{Bacca}
for degenerate conditions, and compare with the results obtained from
the one-pion-exchange
(OPE) approximation to nuclear interactions, which provides the
standard rates for bremsstrahlung in supernova simulations
\citep{HR}. At this level, the transition amplitude is independent of
the center-of-mass momentum and given by the antisymmetrized
interaction $\langle 12 |{\cal T}| 34 \rangle =
\langle 12 |(1-P_{12})V_{\rm NN}| 34 \rangle$ (with exchange operator
$P_{12}$). To be explicit regarding the normalization, the direct and
exchange contributions for the OPE approximation are given by
\be
\langle 12 |(1-P_{12})V^{\rm OPE}_{\rm NN}| 34 \rangle
= - \biggl( \frac{g_A}{2 F_\pi} \biggr)^2 \biggl[ 
\frac{{\bm \sigma}_1 \cdot \kk \, {\bm \sigma}_2 \cdot \kk}{k^2 + m^2_\pi}
- \frac{{\bm \sigma}_1 \cdot \kk' \, {\bm \sigma}_2 \cdot \kk'
+k'^2 ( 1 - {\bm \sigma}_1 \cdot {\bm \sigma}_2)/2}{k^{\prime 2} + m^2_\pi}
\biggr] \,,
\label{OPEamp}
\ee
with pion decay constant $F_\pi = 92.4 \mev$, neutral pion mass $m_\pi
= 134.98 \mev$, and momentum transfers $\kk = \pp_1 - \pp_3$ and $\kk'
= \pp_1 - \pp_4$. For the calculation based on chiral EFT
interactions, we make a partial-wave expansion with the convention
$|{\bf p} \rangle = \sum_{lm} 4\pi \, \rmi^\ell \,
Y^*_{lm}(\widehat{\bf p}) \, |p\rangle |\ell m\rangle$.
After coupling to spin, this leads to
\begin{align}
\Xi(\omega) &= \frac{2}{\sqrt{\pi} (m^* T)^{3/2}} \,
\frac{2\sinh[\omega/(2T)]}{\omega/T} \, \frac{4 \, 
(4 \pi)^2}{3} \, \sum_{\ell \ell' j \tilde{j} L} \, \sum_{m_{s} m'_{s}}
(-1)^{j+\tilde{j}+L} \, \bigl( \, \widehat{j} \, \widehat{\tilde{j}} \,
\widehat{L} \, \bigr)^2 \, \widehat{\ell} \, \widehat{\ell'} \, \nonumber \\
\times &\left\{
\begin{array}{c c c}
\ell & \ell' & L \\
1 & 1 & j \\
\end{array}
\right\}
\left\{
\begin{array}{c c c}
\ell' & \ell & L \\
1 & 1 & \tilde{j} \\
\end{array}
\right\}
\left\{
\begin{array}{c c c}
\ell' & \ell & L \\
\ell & \ell' & 0 \\
\end{array}
\right\} 
\biggl[ {\cal C}_{L (m_{s}-m'_{s})1m'_{s}}^{1m_{s}} \biggr]^{2}
(m_{s}^{2}-m_{s}m'_{s})\nonumber \\
&\times \int_0^\infty dp \, p^2 \, \sqrt{p^2+m^*\omega} 
\, \rme^{-p^2/(m^*T)-\omega/(2T)} \,
\bigl\langle \sqrt{p^2+m^*\omega} \, |V_{\ell \ell'}^{j s=1}| p \bigr\rangle
\bigl\langle \sqrt{p^2+m^*\omega} \, |V_{\ell \ell'}^{\tilde{j} s=1}| p 
\bigr\rangle \,,
\label{pwformula}
\end{align}
with standard notation for the quantum numbers as in \citet{LPS}, and
where the sum is over allowed partial waves with matrix elements
$\langle p |V_{\ell' \ell}^{j s}| p' \rangle$.

\begin{figure}[t]
\begin{center}
\includegraphics[scale=0.45,clip=]{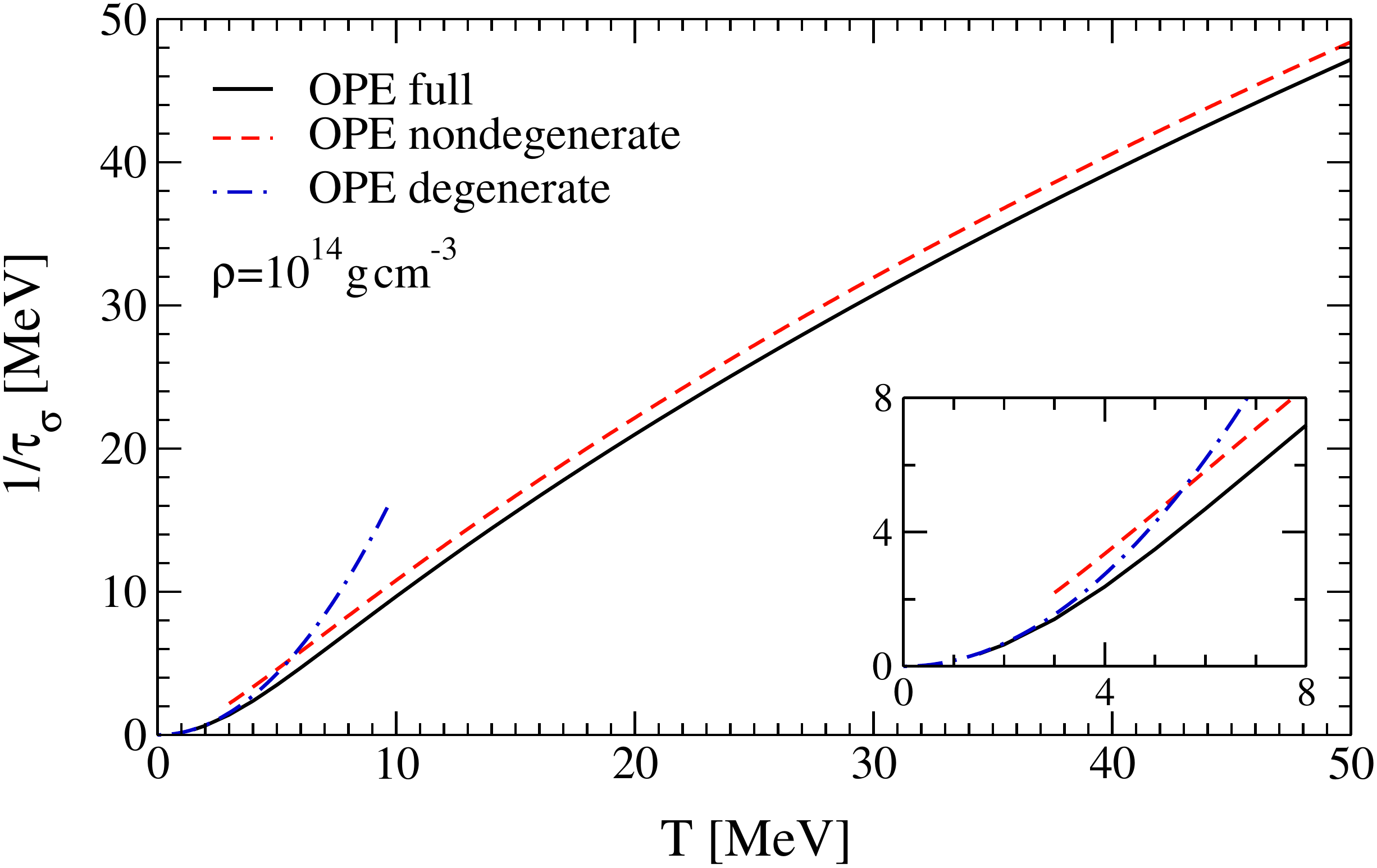} \\[5mm]
\includegraphics[scale=0.45,clip=]{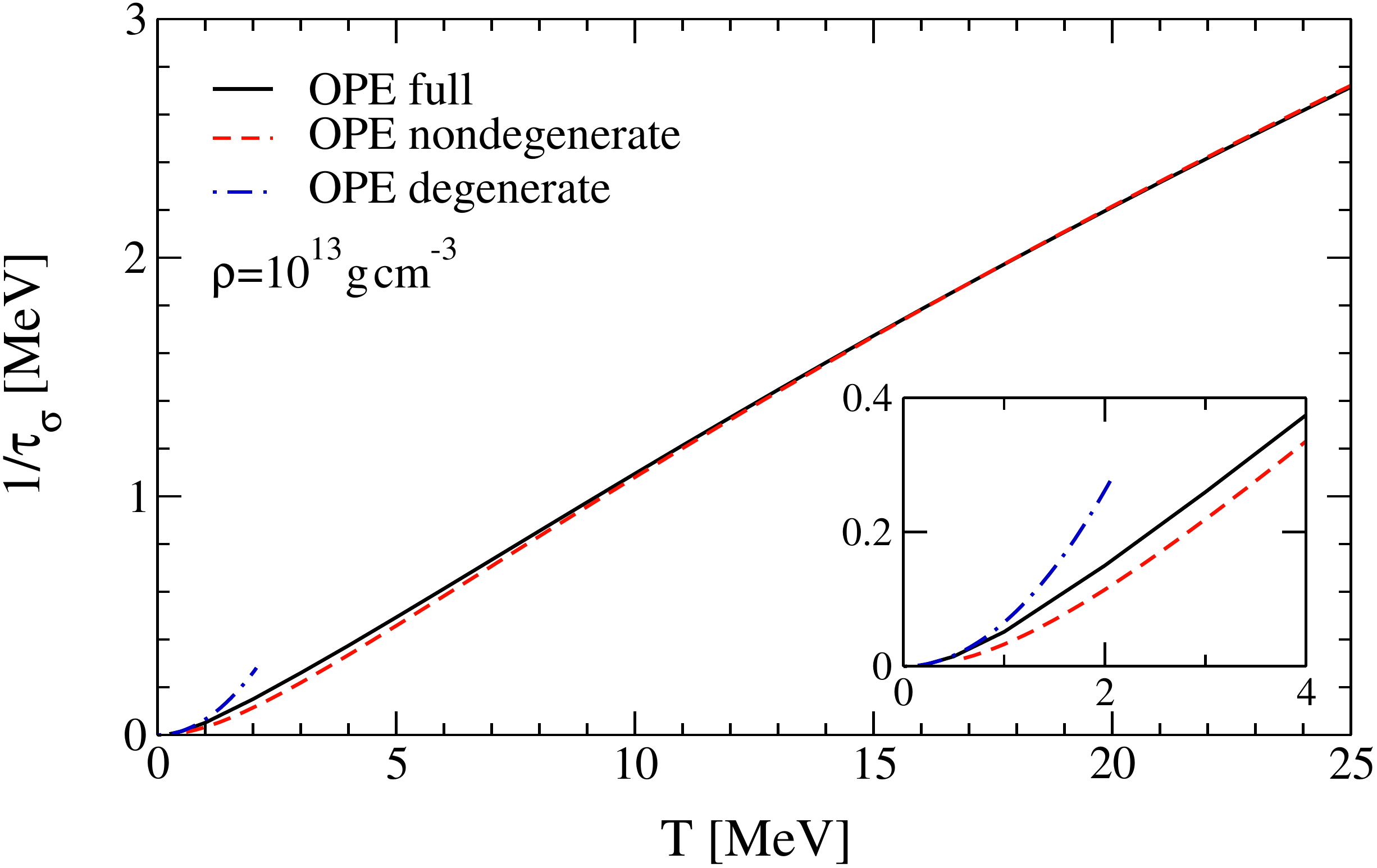}
\end{center}
\caption{Spin relaxation rate $1/\tau_\sigma$ versus temperature $T$
based on the one-pion-exchange (OPE) approximation, comparing the
full (partially degenerate) results to the nondegenerate and
degenerate limits. The top and bottom panels are for density
$\rho = 10^{14} \gcmiq$ and $\rho = 10^{13} \gcmiq$, respectively. The
degenerate curves stop at $T = T_{\rm F}/\pi$. The inset shows the
crossover region from the degenerate to the nondegenerate
regime.\label{rate_regimes}}
\label{rateregimes}
\end{figure}

\begin{figure}[t]
\begin{center}
\includegraphics[scale=0.45,clip=]{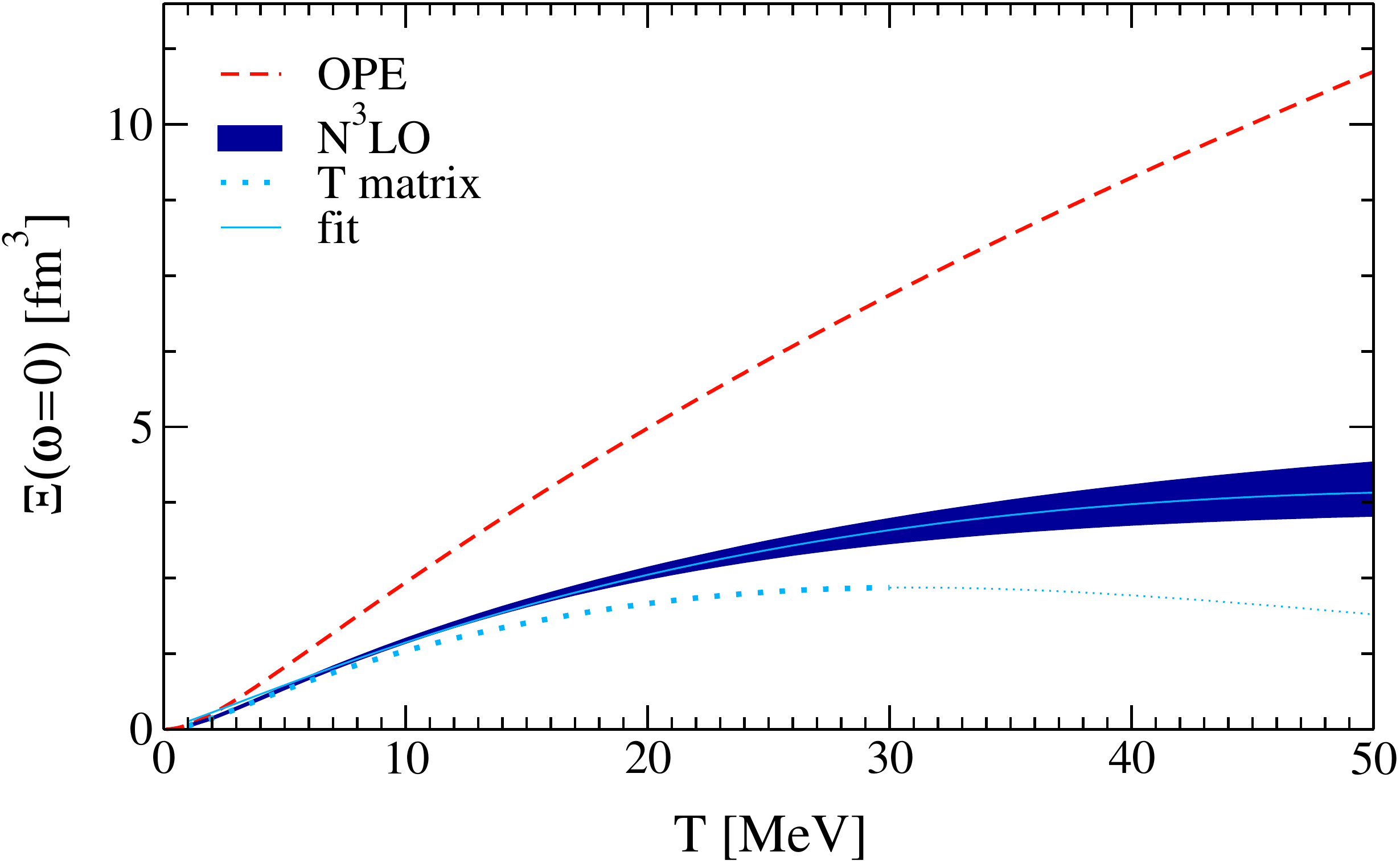}
\end{center}
\caption{$\Xi(\omega=0)$ [Equation~(\ref{defX})] versus temperature
$T$ based on the OPE approximation and from chiral NN interactions at
N$^3$LO, with a fitting curve to the central value of the band.
The N$^3$LO band is due to different NN potentials
employed~\citep{EM,EGM}. In addition, we show results based on the T
matrix for NN scattering using phase shifts and mixing angles from
\citet{nnonline} for $E_{\rm lab} \leqslant 350 \mev$. For temperatures
$T \gtrsim 30 \mev$, the T matrix results are shown by a thin line,
because contributions from higher $E_{\rm lab}$ to
$\Xi(\omega=0)$ can become important.\label{fig:pW}}
\end{figure}

\begin{figure}[t]
\begin{center}
\includegraphics[scale=0.45,clip=]{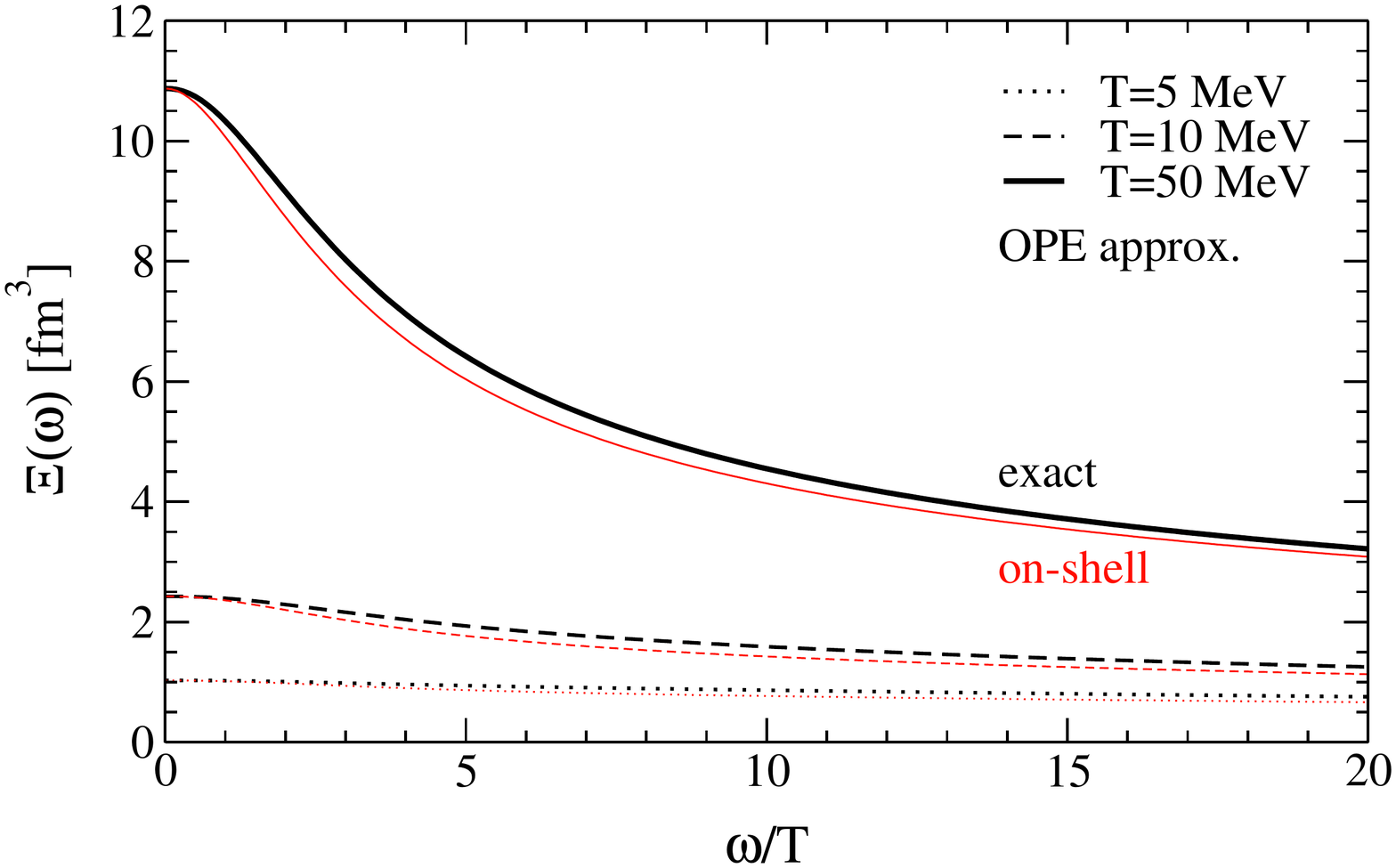}
\end{center}
\caption{$\Xi(\omega)$ [Equation~(\ref{defX})] versus $\omega/T$
for temperatures $T=5,10,50 \mev$. The results are based on the OPE
approximation. Thick (thin) lines are for the exact $\omega$
dependence (on-shell approximation, see text).\label{fig:pW_omega}}
\end{figure}

We begin by comparing results in the OPE approximation allowing for
partial degeneracy with those for the nondegenerate and degenerate
limits. For all results in the following, the neutron effective mass
$m^*$ is taken to be the bare mass $m$.  In Figure~\ref{rateregimes},
we show the spin relaxation rate as a function of temperature for
densities $\rho = 10^{14} \gcmiq$ (top panel) and $\rho = 10^{13}
\gcmiq$ (bottom panel). The full (partially degenerate) results are
based on numerical integrations of
Equation~(\ref{tausigmafinal2}). The spin relaxation rate in the
degenerate limit \citep{LPS} is shown for $T \leqslant T_{\rm F}/\pi$
and the nondegenerate limit is given by
Equation~(\ref{tausigmafinalnondegen}). Our results clearly
demonstrate that the nondegenerate limit is a good approximation to
the full results down to temperatures well below the Fermi
temperature, $T_{\rm F} = 30 \mev$ at a density $\rho = 10^{14}
\gcmiq$, and $T_{\rm F} = 6.5 \mev$ at $\rho = 10^{13} \gcmiq$. In the
nondegenerate case, the spin relaxation rate is linear in the density,
and information about the temperature dependence is contained in the
quantity $\Xi(\omega)$, which is plotted in Figure~\ref{fig:pW} for
$\omega=0$. As expected from the spin relaxation rate in the
degenerate limit \citep{Bacca}, results based on chiral NN
interactions at N$^3$LO are typically a factor of two smaller than
those obtained using the standard OPE approximation. As in
\citet{Bacca}, we consider different N$^3$LO NN
potentials~\citep{EM,EGM} and the resulting band gives a range of
uncertainty at this level. We also give a fit to the central value
of the N$^3$LO band:
\be
\Xi(\omega=0) = 0.133 \, \biggl( \frac{T}{\rm MeV} \biggr)^{1.12}
e^{-0.02 \, (T/{\rm MeV})} \,.
\label{fit}
\ee
Since at typical frequencies and temperatures, the $\omega$ dependence
is weak (see Fig.~\ref{fig:pW_omega}), we recommend substituting the
$\omega=0$ fit, Equation~(\ref{fit}), to calculate the spin relaxation rate,
Equation~(\ref{tauX}). This can be combined with Equations~(\ref{structspin})
and~(\ref{susc}) for the spin dynamical structure factor to explore
our rates for neutrons in astrophysical simulations.

In addition, we compare the N$^3$LO results to those based on the T
matrix for NN scattering, which has been used for degenerate neutrons
in \citet{Hanhart}. For $\omega=0$, the
partial-wave potential matrix elements $\bigl\langle p |V_{\ell
\ell'}^{j s=1}| p \bigr\rangle \bigl\langle p |V_{\ell \ell'}^{\tilde{j}
s=1}| p \bigr\rangle$ in Equation~(\ref{pwformula}) are replaced by
on-shell T matrices, $\bigl\langle p |T_{\ell \ell'}^{j s=1}| p \bigr\rangle^* 
\bigl\langle p |T_{\ell \ell'}^{\tilde{j} s=1}| p \bigr\rangle$
(note the complex conjugate), which are give in terms of scattering
phase shifts and mixing angles [see, e.g., \citet{BrownJackson}]. 
For NN scattering, we take the phase shifts and mixing angles from
\citet{nnonline} for $E_{\rm lab} \leqslant 350 \mev$, and note that
contributions from higher $E_{\rm lab}=2 \, p^2/m$ can become
important for temperatures $T \gtrsim 30 \mev$. For temperatures
$T < 30 \mev$, the T matrix results are similar, but somewhat lower
compared to those from chiral NN potentials at N$^3$LO. This
indicates that noncentral neutron-neutron interactions may be
perturbative in chiral EFT for the energies relevant to the spin
relaxation rate (where low energies are also suppressed by one
power in momentum $p$ at $\omega=0$).

We now consider how good the expression for the spin relaxation rate
in the nondegenerate limit is under relevant conditions encountered in
simulations of core-collapse supernovae. As representative of
conditions during outflow, based on Figures~\ref{fig:conditions15}
and~\ref{fig:conditions40}, we take for values of the density ranging
between $\rho_0/100$ and $\rho_0$ the temperature given by
Equation~(\ref{SNconditions}).

\begin{figure}[t]
\begin{center}
\includegraphics[scale=0.45,clip=]{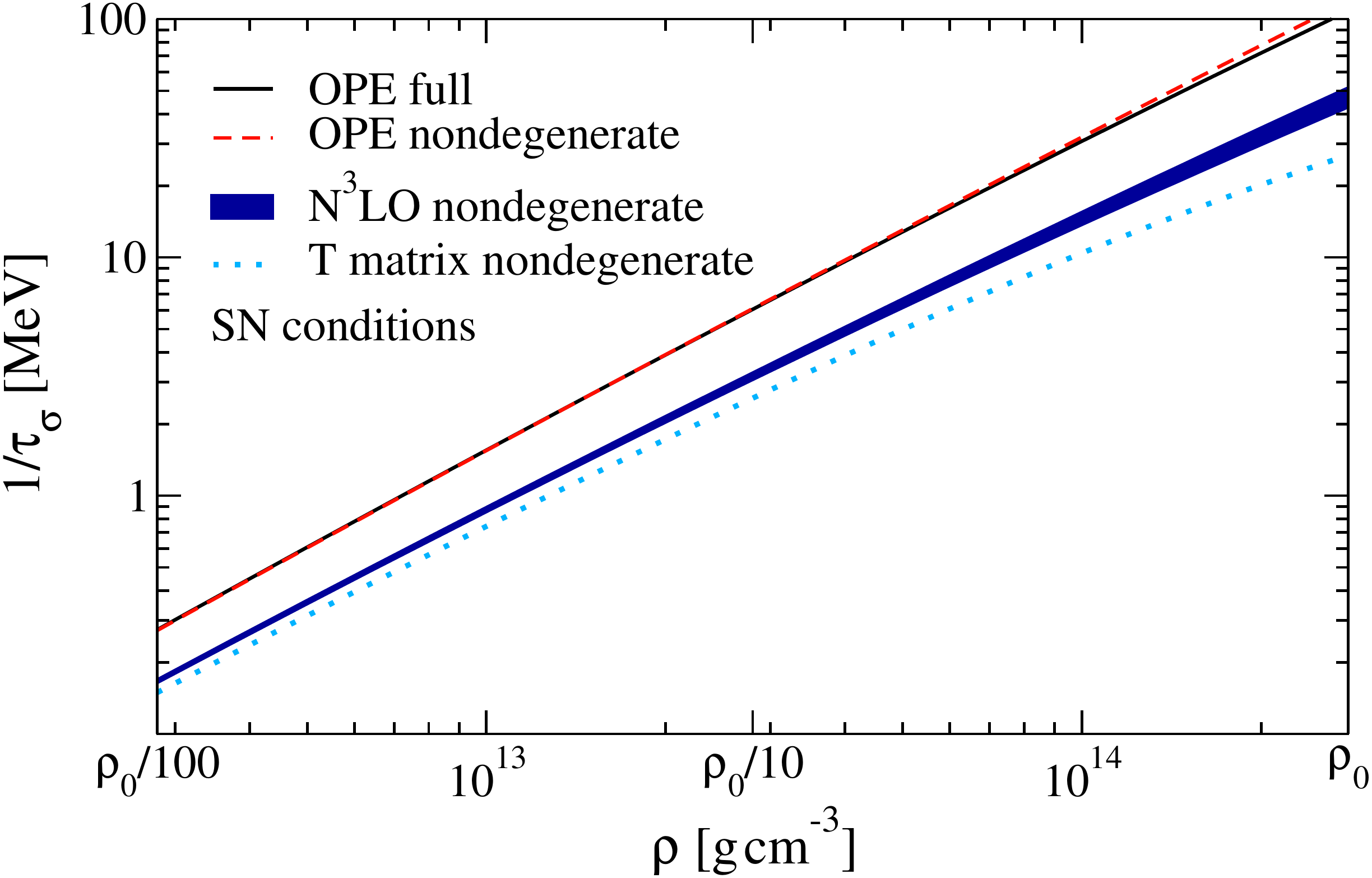}
\end{center}
\caption{Spin relaxation rate $1/\tau_\sigma$ versus density $\rho$ 
for conditions typical of the outflow phase in supernova simulations
[given by Equation~(\ref{SNconditions})]. The full (partially
degenerate) results are compared to the nondegenerate limit for the
OPE approximation. In addition, we show the spin relaxation rate
based on chiral NN interactions at N$^3$LO and based on the T
matrix for NN scattering. Note that the density
dependence is linear in the nondegenerate limit [see
Equation~(\ref{tauX})].\label{rate_SN}}
\end{figure}

In Figure~\ref{rate_SN}, the full (partially degenerate) results are
compared to the nondegenerate limit for these conditions. This figure
demonstrates that results for the nondegenerate limit are an excellent
approximation for matter in supernova simulations at subnuclear
densities. Moreover, for the broad density range considered, the
results based on N$^3$LO NN potentials are similar to those based on
the NN T matrix, and both lead to spin relaxation rates that are a
factor two or more smaller compared to the OPE approximation.

\begin{figure}[t]
\begin{center}
\includegraphics[scale=0.45,clip=]{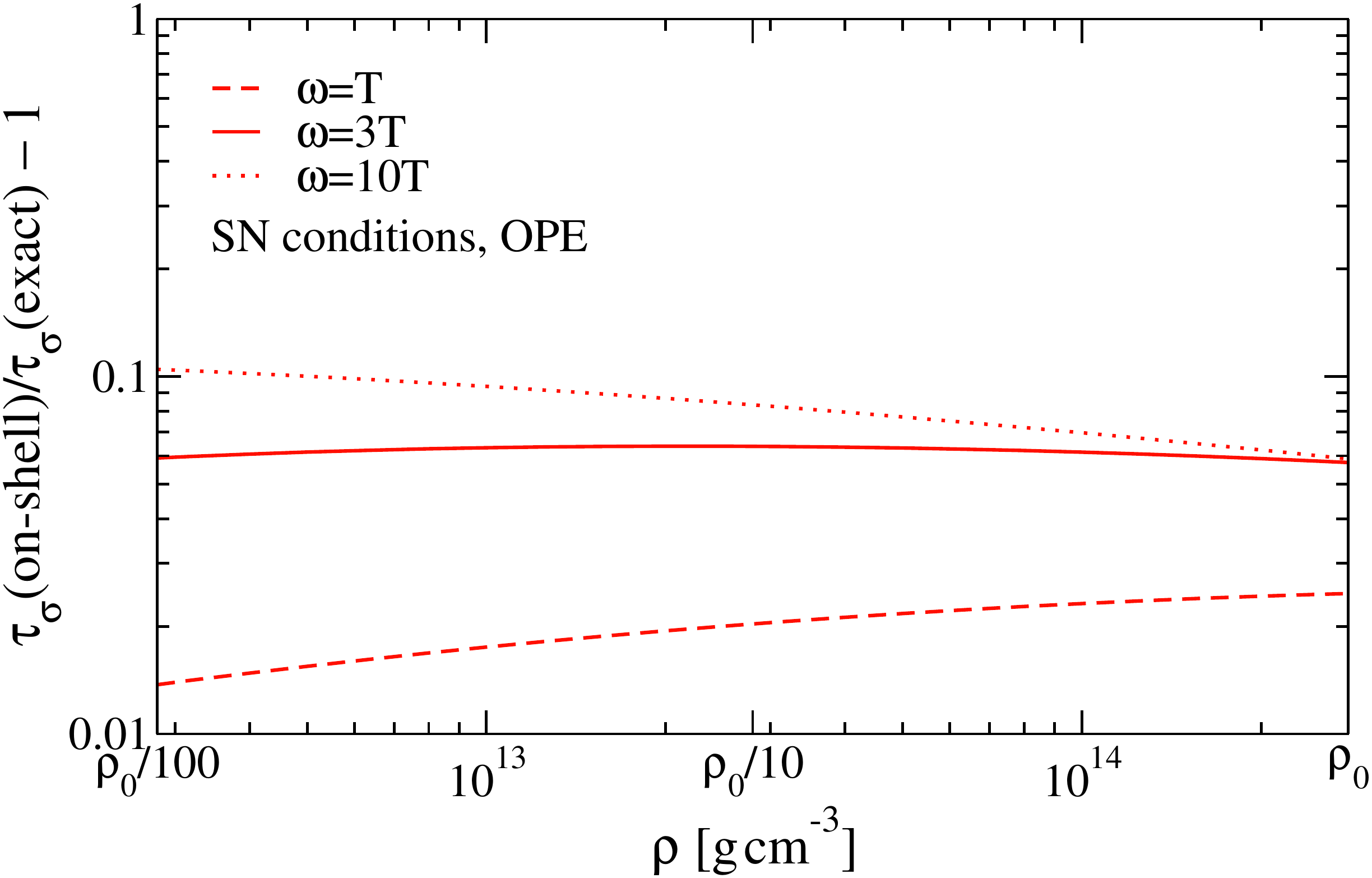}
\end{center}
\caption{Relative difference, $\tau_\sigma(\text{on-shell})
/\tau_\sigma(\text{exact})-1$, versus density $\rho$ based on the
OPE approximation, comparing the exact $\omega$ dependence to the
on-shell approximation (see text) for $\omega=T, 3T$, and $10T$. The
comparison is for conditions typical of the outflow phase in
supernova simulations [given by Equation~(\ref{SNconditions})].
\label{reldiff}}
\end{figure}

Next, we study the dependence of the spin relaxation rate in the
nondegenerate limit on frequency. When $\omega \neq 0$, the matrix
elements that enter the rate are off-shell for $p \neq p' =
\sqrt{p^2+m^*\omega}$ [see Equation~(\ref{pwformula})], because the
energies of the particles in the initial and final states differ by
$\omega$. Figure~\ref{fig:pW_omega} shows $\Xi(\omega)$ as a function
of $\omega/T$ for the OPE interaction for three different
temperatures, and compares the exact $\omega$ dependence with results in
the on-shell approximation, where the partial-wave matrix elements are
evaluated at a mean energy
$V\bigl(\sqrt{p^2+m^*\omega/2},\sqrt{p^2+m^*\omega/2} \bigr)$. The
differences are found to be small, because the momentum dependence of
the interaction matrix elements varies on a scale of the order of the
pion mass (and higher momenta for chiral EFT interactions), which is
large compared to the energy transfer, which is of the order of the
temperature.  This conclusion is reinforced by the plots of the
relative difference of the spin relaxation rate calculated with the
exact $\omega$ dependence compared to the on-shell approximation in
Figure~\ref{reldiff} for typical conditions in supernova simulations.

\begin{figure}[t]
\begin{center}
\includegraphics[scale=0.45,clip=]{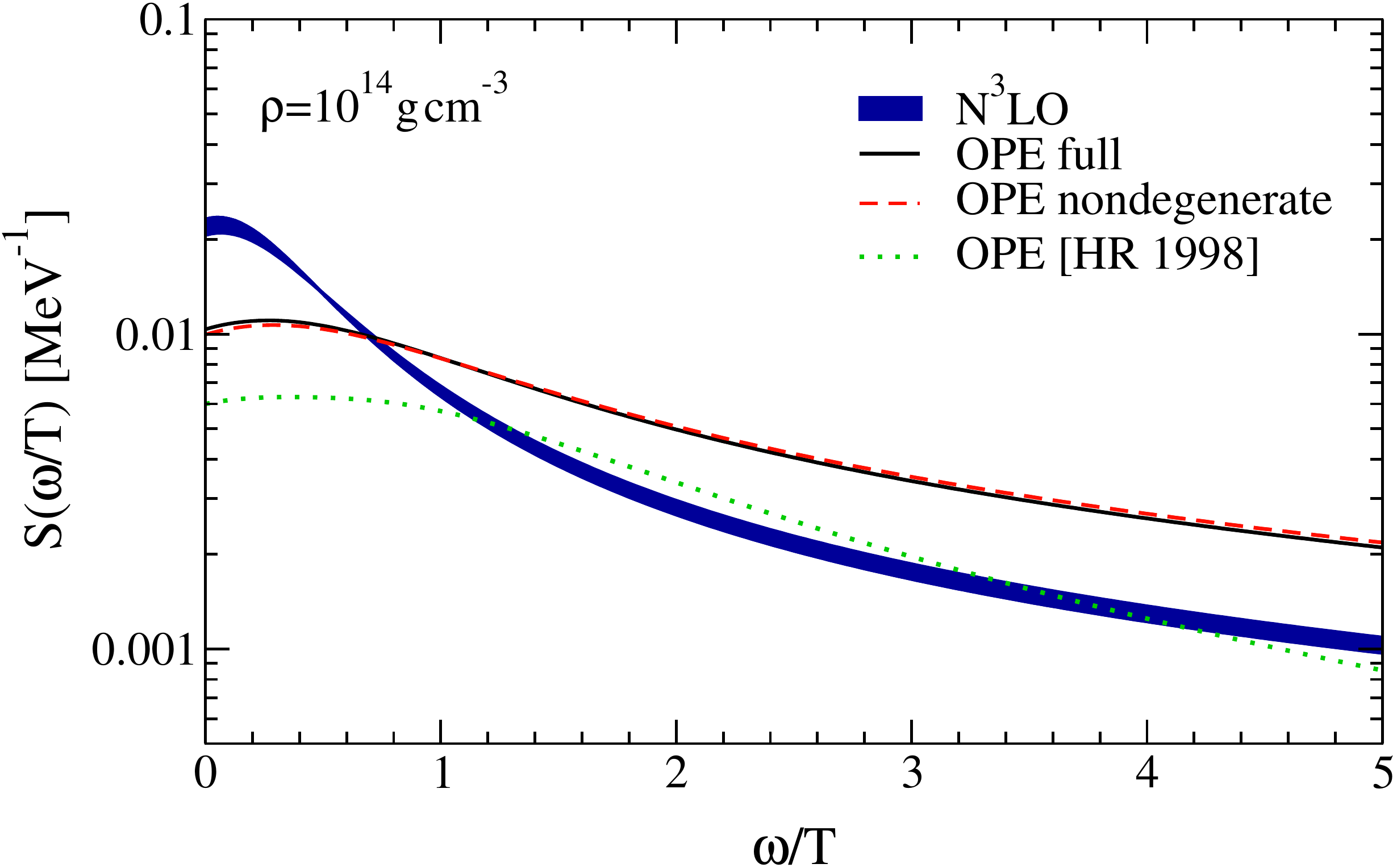} \\[5mm]
\includegraphics[scale=0.45,clip=]{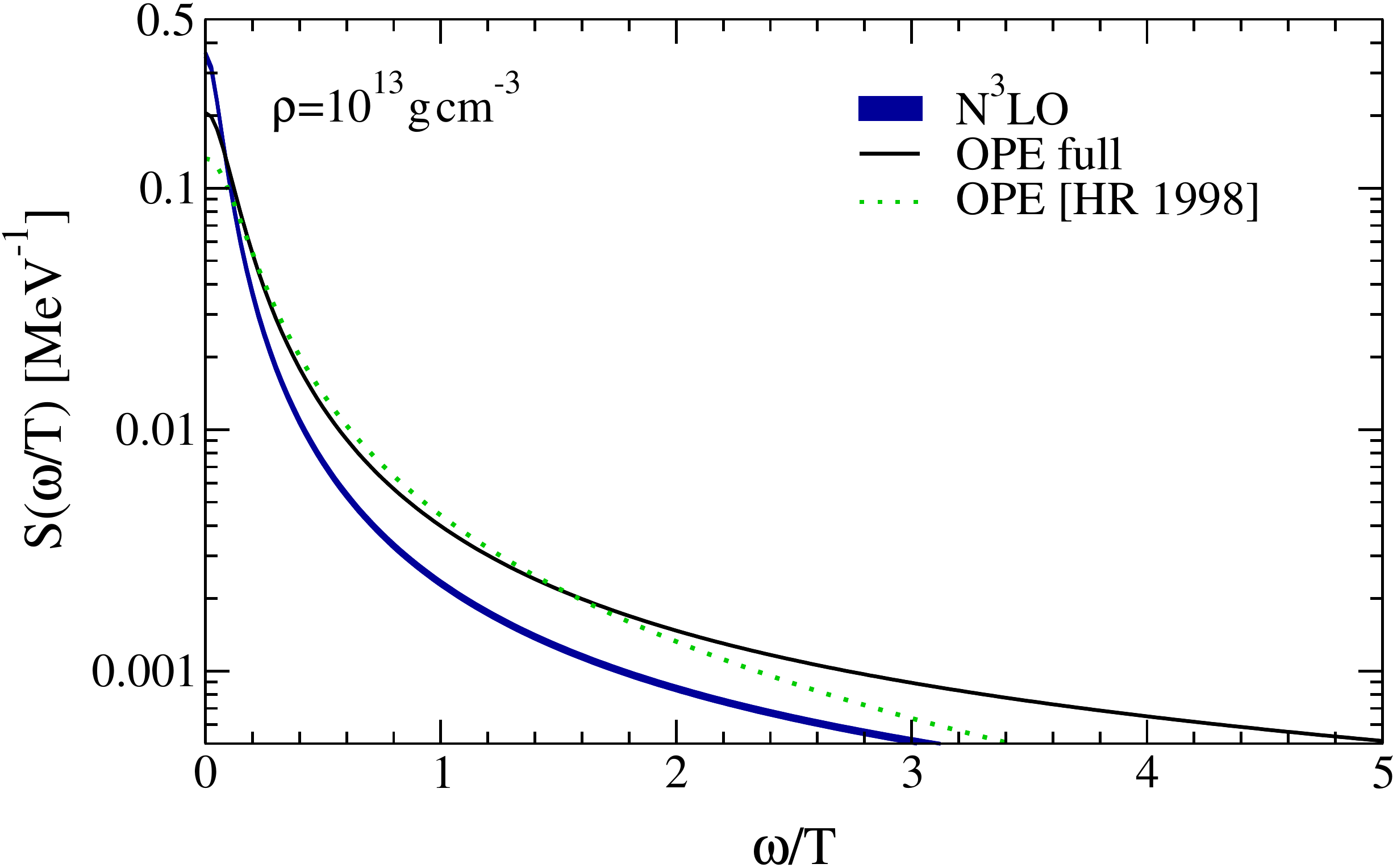}
\end{center}
\caption{Spin dynamical structure factor $S(\omega/T)$ at long
wavelength $q=0$ as a function of energy transfer $\omega$ divided
by the temperature $T$ for densities $10^{14} \gcmiq$ (upper panel)
and $10^{13} \gcmiq$ (lower panel). The corresponding temperatures
are given by Equation~(\ref{SNconditions}). The blue bands show our results
based on chiral NN interactions at N$^3$LO, while the solid lines 
show results for the OPE approximation. For $\rho = 10^{14} \gcmiq$,
we also compare the nondegenerate approximation (dashed line) with
the full OPE results, while at $10^{13} \gcmiq$ the results for the 
nondegenerate approximation are indistinguishable from the full 
calculation on the scale of the figure. The dotted lines show the
OPE results of \citet{HR} [HR 1998] as implemented in the Basel
code for core-collapse supernova simulations \citep{Fischer2012}.
\label{structurefactorplot}}
\end{figure}

For astrophysical applications, the quantity that enters the rates of
neutrino processes is the spin dynamical structure factor which is
related to the relaxation rate by Equations~(\ref{structspin})
and~(\ref{susc2}). In Figure~\ref{structurefactorplot}, we plot the
structure factor for two different densities, $10^{13} \gcmiq$ and
$10^{14} \gcmiq$ [with corresponding temperatures given by
Equation~(\ref{SNconditions})]. The blue bands show the results for
the N$^3$LO NN potentials and the solid lines those for the OPE
approximation calculated within the formalism of this paper. As one
would expect from the fact that the width of the peak in the structure
factor is proportional to the relaxation rate and its height is
inversely proportional to the rate, the results for the N$^3$LO
interactions have a higher peak value and a narrower width compared
with those for the OPE approximation. As representative of earlier
work, we show as dotted lines results for the OPE calculations of
\citet{HR} as implemented in the Basel code for core-collapse
supernova simulations \citep{Fischer2012}. For a density of $10^{14}
\gcmiq$, the \citet{HR} results lie 30-60\% below our OPE results,
while at $10^{13} \gcmiq$, the difference is generally less. The
difference between the OPE results may be a consequence of imposing a
normalization condition in \citet{HR}, while in our approach the width
is obtained consistently by solving the Boltzmann equation. For the
comparison in Figure~\ref{structurefactorplot}, we have neglected the
frequency dependence of the relaxation rate, which is weak at these
temperatures and would increas the structure factor with increasing
$\omega/T$ (see Figure~\ref{fig:pW_omega}), as well as Fermi liquid
corrections [$g_0=0$ in Equation~(\ref{susc2})], which are not
included in \citet{HR}.

In scattering, the energy transfer to or from neutrinos is comparable
to the width of the peak in the structure factor, and thus one sees
that the energy transfer under realistic supernova conditions can be
comparable to the temperature. It is important in future simulations
to allow for this energy transfer. In addition, the rate of neutrino
pair bremsstralung is roughly proportional to the spin relaxation
rate, and therefore the N$^3$LO results predict a lower rate for this
process than does the OPE approximation. This is consistent with what
has previously been found for degenerate conditions
\citep{Bacca}. More comprehensive comparisons of predictions for rates
of neutrino processes will be presented in future work that will
extend the present calculations to take into account the effect of protons.

\section{Concluding remarks}
\label{summary}

We have analyzed simulations of core-collapse supernovae and have
shown that in the post-bounce phase the relevant conditions for
neutrino processes are partially degenerate or nondegenerate. We then
developed a formalism for calculating neutrino rates in strongly
interacting matter by generalizing to the partially degenerate regime
the approach used for degenerate matter in \citet{LPS}. The resulting
spin dynamical structure factor takes into account both mean-field
effects and collisions between excitations in neutron matter. We then
calculated the spin relaxation rate, a key ingredient in the structure
factor which enters expressions for the rates of neutrino processes.
This was done at two levels of NN interactions, the OPE approximation,
which is commonly used in simulations \citep{HR}, and chiral EFT
interactions. We have found that chiral NN interactions lead to a
reduction of the spin relaxation rate typically by a factor of two for
a broad range of conditions. This reduction is similar to what
previously has been found in the degenerate regime
\citep{Bacca}. We have also found that our OPE rate, where the width
is obtained consistently by solving the Boltzmann equation, differs
from the OPE results of \citet{HR}. This may be a consequence of 
the imposition of a normalization condition in \citet{HR}.
Moreover, our results demonstrate that the
nondegenerate limit is an excellent approximation for the conditions
encountered in the post-bounce phase of matter at subnuclear densities
in supernova simulations.

Future directions include the study of many-body contributions and of
many-body forces and electroweak currents in chiral EFT, and the
extension of the calculations to mixtures of neutrons and protons,
where the central parts of nuclear interactions can cause relaxation
of the axial charge, because of the different axial charges of the
neutron and proton. These extensions will be greatly simplified by the
finding that the nondegenerate limit provides an excellent
approximation for the relevant supernova conditions.

Throughout the paper we have assumed that the basic coupling of the
weak neutral field (that of the $Z$ boson) to nucleons is via a
one-body operator. However, the spin relaxation effect that we have
considered amounts to coupling a single quasiparticle-quasihole pair
to two quasiparticle-quasihole pairs. In other words, strong
interactions have generated a two-body contribution to the weak charge
operator. However, not all two-body contributions to the operator are
included in calculating relaxation effects by this procedure. The
approximation we have made includes only contributions with a single
quasiparticle in an intermediate state, which are enhanced by
$1/\omega$. Consideration of other two-body contributions to the weak
charge operator is left for future studies.

\begin{acknowledgments}
We thank the Niels Bohr International Academy and NORDITA for their
hospitality. This collaboration was facilitated by the participation
of a number of authors in the MICRA2009 workshop in Copenhagen,
which was supported in part by the ESF AstroSim and CompStar
networks. A.P.~thanks Tobias Fischer for providing simulations
data. This work was supported by NSERC and the NRC
Canada, by the Helmholtz Alliance Program of the Helmholtz
Association, contract HA216/EMMI ``Extremes of Density and
Temperature: Cosmic Matter in the Laboratory'', and by the DFG
through Grant SFB 634.
\end{acknowledgments}

\end{document}